\documentclass{emulateapj}
\usepackage{epsfig}
\usepackage{amsmath}
\usepackage{amssymb}
\usepackage{natbib}
\usepackage{setspace}
\usepackage{graphicx}

\def \bhalias {BH14}

\begin{document}
\title{How galactic environment regulates star formation }
\author{Sharon E. Meidt\altaffilmark{1}}

\altaffiltext{1}{Max-Planck-Institut f\"ur Astronomie / K\"{o}nigstuhl 17 D-69117 Heidelberg, Germany}

\date{\today}
\begin{abstract}
In a new simple model I reconcile two contradictory views on the factors that determine the rate at which molecular clouds form stars -- internal structure vs. external, environmental influences -- providing a unified picture for the regulation of star formation in galaxies. In the presence of external pressure, the pressure gradient set up within a self-gravitating isothermal cloud leads to a non-uniform density distribution.  
Thus the local environment of a cloud influences its internal structure.  In the simple equilibrium model, the fraction of gas at high density in the cloud interior is determined simply by the cloud surface density, which is itself inherited from the pressure in the immediate surroundings.  
This idea is tested using measurements of the properties of local clouds, which are found to show  remarkable agreement with the simple equilibrium model.  The model also naturally predicts the star formation relation observed on cloud scales and, at the same time, provides a mapping between this relation and the closer-to-linear molecular star formation relation measured on larger scales in galaxies.  The key is that pressure regulates not only the molecular content of the ISM but also the cloud surface density.  I provide a straightforward prescription for the pressure regulation of star formation that can be directly implemented in numerical models.  
Predictions for the dense gas fraction and star formation efficiency measured on large-scales within galaxies are also presented, establishing the basis for a new picture of star formation regulated by galactic environment.  
\end{abstract}
\section{Introduction\label{sec:intro}}
Our view of the nature of star formation has undergone a dramatic change in recent years, following ground-breaking observations of the seeds of star-formation, giant molecular clouds (GMCs), both within our own galaxy and in the extragalactic context.  
Surveys of clouds mapped with unprecedented detail in dust extinction, dust emission and molecular line emission in our own Milky Way have revealed a complex internal cloud structure encompassing clumps and cores within a filamentary web (e.g. \citealt{andre}; \citealt{molinari}).  
These clouds can be linked directly with infrared catalogs of young-stellar objects (YSOs) that provide a direct, unbiased inventory of current star formation (e.g. Evans et al. 2003, 2009 and \citealt{dunham}; see \citealt{ladall09}; \citealt{heiderman}).  

Meanwhile, the first extragalactic observations covering large areas in nearby galaxies at cloud-scale resolution have placed clouds in the context of their dynamical environment (\citealt{rosol}; \citealt{hughesLMC}; \citealt{koda}; \citealt{rebolledo}; \citealt{donovanmeyer}; \citealt{schinnerer}; \citealt{colombo2014a}).  
These suggest a departure from the standard view, namely that clouds are decoupled from their surroundings, and reveal instead that clouds inherit their properties from their environment (\citealt{hughesII}; \citealt{colombo2014a}).  
Paired with multi-wavelength data sets, these observations have begun to establish a connection between gas stability and cloud properties and the global pattern of star formation in galaxies (\citealt{meidt}; \citealt{hughesI}).  

At the intersection of these two lines of investigation (cloud studies in either the MW or external galaxies), one recurring question continues to emerge: do clouds obey the same empirical scaling relation between star formation rate (SFR) and gas surface density followed on large scales in galaxies?  
The relation -- the Kennicutt-Schmidt (KS; \citealt{schmidt}; \citealt{kennKS}) law -- holds over many orders of magnitude in gas content and has been interpreted as encapsulating 
 the fundamental process of star formation at the cloud scale.  Evidence at or below the scale of individual clouds, however, is more ambiguous as to the nature and origin of the relation.  In external galaxies, the KS law appears to break down with increasingly high spatial resolution.  
 Several explanations have been proposed, including discreteness effects \citep{feldmann}, temporal and spatial offsets between gas and young stars (e.g. \citealt{schruba}; \citealt{kruijssen}), and dynamical influences on gas stability \citep{meidt}.  
 Studies of the local cloud population also even suggest that no star formation relation exists among individual clouds (\citealt{heiderman}; \citealt{lada2013}).  

Clouds do appear to internally obey the Kennicutt-Schmidt relation (e.g. \citealt{ladall09}), however, in that star formation is observed to be most directly associated with gas at the highest densities.  
This has been suggested by numerous studies of local clouds, in which dense gas is traced with a variety of techniques (\citealt{onishi}; \citealt{enoch}; \citealt{johnstone}; \citealt{andre}; \citealt{li97}; \citealt{lada92}; \citealt{wu05}; \citealt{heiderman}).  
A direct link between dense gas content and star formation in nearby molecular clouds was established by \citet{ladall10} who found a correlation between the number of YSOs and the dense gas in clouds above an extinction of $A_K$ = 0.8, corresponding to $A_V$=8.   

Recently \citet{evansHeiderman} have confirmed this picture, showing that the mass of dense gas measured in local clouds is more closely linked to the SFR than any other measurable cloud property, including surface density over free-fall time or crossing time.  
They therefore conclude that there is no fundamental, universal relation between $\Sigma_{SFR}$ and $\Sigma_{gas}$, highlighting instead that the SFR per unit dense gas mass is roughly independent of the cloud dense gas mass.  

A new picture is therefore emerging, wherein star formation depends only on the internal structure of the cloud, not on any global cloud property.  
Yet such a conclusion is difficult to reconcile with results on larger scales in external galaxies, which suggest that the environment in which a cloud lives determines its properties and its ability to collapse and form stars (\citealt{hughesII}; \citealt{meidt}) and with theories of self-regulated star formation (\citealt{ostrikerShetty}; \citealt{hopkinsReg}).  

This paper aims to unite these two views by examining what sets the dense gas fraction in clouds.  In section \ref{sec:model} I introduce the model of a pressurized cloud, which describes how local environment determines the surface density and initial internal structure of clouds. 
The model predicts a link between the dense gas mass fraction (DGMF) and the cloud surface density (section \ref{sec:equilStructure}), which is ideally tested against real clouds, given two measurements at two different densities per cloud.    
Section \ref{sec:tests} reexamines the cloud sample studied by \citet{battistiHeyer}~for consistency with the model and reveals good agreement with the predicted trend.  

Section \ref{sec:SFimplications} explores the implications of the pressurized cloud model, first introducing a new interpretation for the observed cloud-scale relation between star formation rate and cloud surface density (section \ref{sec:localSF}).  
The new model is considered in relation to local MW clouds studied in extinction by \citet{heiderman}, which provide compelling evidence in favor of the influence of pressure as formulated here (section \ref{sec:HeidTest}). 
Then Section \ref{sec:globalSF} describes how the pressurized cloud model provides a natural mapping between the super-linear cloud-scale relation and the extragalactic linear star formation relation measured on large-scales in nearby galaxies.  
The key is regulation of cloud surface density and molecular content of the ISM by pressure.  Leveraging this idea, section \ref{sec:predictionsDense} introduces predictions for trends in the dense gas fraction and dense gas star formation efficiency in galaxies, which can be tested with observations of dense gas tracers accessible, i.e.,  with ALMA.  
Section \ref{sec:conclusions} brings the paper to a close with a summary of the key aspects of the model and the results of the preliminary testing performed here.  

\section{Pressurized cloud model}
\label{sec:model}
\subsection{The motivation}
\label{sec:motivation}
One of the resounding elements in models of star formation is that molecular clouds have `universal' properties and exist in a state of virial equilibrium, according to the balance observed between their gravitational and turbulent (kinetic) energies and the overall agreement of measured cloud properties with standard scaling relations (i.e. Larson's laws, which follow from virialization; \citealt{larson81}, \citealt{mckee92}, \citealt{heyer09}, \citealt{bolatto} and references therein).  This idea has met with some skepticism, however, in light of the challenges of determining properties like mass, radius and velocity dispersion with observation (and in simulations), which makes it difficult to reliably assess the balance of energies within a cloud.   The inferred energy balance also tends to ignore factors like magnetic fields, surface terms and projection effects (\citealt{ballesteros06}; \citealt{dib07}; \citealt{shetty}).   

At present, a growing number of studies associate departures from a balance between cloud kinetic and gravitational potential energies with the presence of external pressure (\citealt{keto}; \citealt{elmegreen}; \citealt{bertoldimckee}; \citealt{heyerbrunt}; \citealt{lada2008}; \citealt{field}; \citealt{colombo2014a}) or with a genuine departure from equilibrium as might be characteristic of dynamically evolving structures (e.g. Dobbs et al. 2009).  The idea that clouds experience a non-negligible external pressure would offer a compelling explanation for the observed variation in cloud properties with dynamical environment (arm vs. interarm) recently observed in the grand-design spiral galaxy M51 \citep{colombo2014a}, and from the Galactic center to the disk \citep{sanders}, as well as from galaxy to galaxy \citep{hughesII}.  In such a picture, clouds inherit their properties from their environment given the coupling of clouds to their surroundings via external pressure.  This is notably in contrast to the `universal' cloud picture invoked by modern theories (\citealt{krumholz}; \citealt{ostriker}; \citealt{ostrikerShetty}).  

Support for the idea that clouds are coupled to their surroundings has been found in a recent test of the standard `universal' cloud paradigm, in which \citet{hughesII} show that the ensemble of clouds in M51, as well as in the lower mass galaxies M33 and the LMC (among other galaxies), exhibit a near balance between internal and external pressure.   The external pressure $P_{ext}$ on clouds is measured via hydrostatic equilibrium, including the weight of the stars and gas, and the internal, kinetic pressure is measured as $P_{int}=\rho\sigma^2$, where $\rho$ is the volume density of the (spherical) cloud and $\sigma$ is the measured velocity dispersion. 
This similarity in pressures is in contrast to earlier considerations (e.g. \citealt{blitz}), which take the large density contrast between clouds and the surrounding gas disk as evidence that they are decoupled from their environment.  
However, this argument overlooks the contribution of the stellar disk to the hydrostatic mid-plane pressure, which \citet{hughesII} find is responsible for equal, if not more, external pressure on clouds in the disks of normal star-forming galaxies than exerted by the neutral ISM alone.  
It also ignores a component of the molecular medium outside clouds traced by diffuse CO emission, and which appears to reside in a puffy, vertically extended disk, as observed in M51 \citep{pety} and inferred in several more nearby galaxies \citep{caldu}. 

Although the detailed balance between internal and external pressures likely varies from galaxy to galaxy (depending on the properties of the clouds and the gaseous and stellar disks), even in normal galaxies it thus appears that clouds may not be quite so strongly decoupled from their surroundings as once thought.  This can explain why, unlike the clouds analyzed by \citet{bolatto} that seem to support the `universal' cloud picture, clouds from across a greater diversity of environments fail to exhibit the expected `universal' cloud properties and constant surface density, in particular \citep{hughesII}. 

\subsection{The basic picture}
In light of this evidence, this paper considers a model in which all clouds (even in the disks of nearby galaxies, not just in the case of gas-rich starbursts) are pressurized at their surface, although not necessarily pressure-confined (i.e. with $P_{ext}$$\gtrsim$$P_{int}$).  
The equilibrium density distributions in such clouds establish a link in internal cloud structure over a range of spatial scales.    
The goal here is to relate cloud-scale structure, on the order of tens of parsecs, to the internal structure down to 0.1-1 pc, within which high densities corresponding to $A_v$$>$8 are reached. 
The material at these high densities forms the filaments, clumps and cores observed at even smaller scales where the majority of young stars are observed (i.e. \citealt{ladall10}; \citealt{evansHeiderman}). 

As developed more below, the model consists of a cloud subject to external pressure under the simplest of assumptions, namely a spherical geometry and no magnetic pressure. 
The cloud of mass $M$ is further assumed to be isothermal with equation of state $P=\rho c_s^2$, but an additional non-thermal (turbulent) component that can act to support the cloud is also allowed so that, more generally $P=\rho\sigma^2$ with velocity dispersion $\sigma$ representing an effective sound speed that includes isotropic turbulent motion.   
The additional simplifying assumption that $\sigma$ is uniform within the cloud is also made.    
This assumption is arguably valid across the scales in the range 1-20 pc we are considering here (cf. \citealt{myers}; as discussed in the Appendix and considered in Figure \ref{fig:sigplot} there) in which resides the bulk of the cloud mass.\footnote{As described in what follows, observed density profiles are consistent with $\rho$$\propto R^{-n}$ with $n\lesssim$2). Very little mass is situated at small radii (at the highest densities).  }

For this type of cloud, the equation for hydrostatic equilibrium is
\begin{equation}
\frac{dP}{dR}=-\frac{\rho GM}{R^2}\label{eqn:hydro}
\end{equation}
where the surface pressure is equal to the external pressure at the outer cloud boundary.    
The external pressure exerted on this so-called `Bonnor-Ebert' cloud establishes a radial pressure gradient within the cloud, leading to a non-uniform density distribution typically parameterized as 
\begin{equation}
\rho=\frac{\rho_c a^2}{R^2+a^2}
\end{equation}
which varies as $R^{-2}$ at large distances from the characteristic flat core with density $\rho_c$ and size 
\begin{equation}
a=k\frac{\sigma}{\sqrt{G\rho_c}}
\end{equation}
written in terms of the Jeans length, with velocity dispersion $\sigma$ and constant of proportionality $k$ (i.e. \citealt{dapp}).  

The resulting equilibrium density profile, described more in the next section, provides the standard against which countless observations of MW clouds have been compared.   
This profile is notoriously difficult to distinguish from collapse, however, at least based on morphology alone.  
This issue will be addressed later in $\S$ \ref{sec:collapse}.  Here it should be emphasized that, with the benefit of kinematic information, $R^{-2}$ profiles appear to be associated with quasi-static objects, at least on core scales (see \citealt{ketoCR}).  
Indeed, this is more the type object envisioned here than a strictly static object in prolonged equilibrium.  Such a picture now regularly emerges from numerical simulations of GMC formation and evolution and extragalactic molecular cloud surveys, in which clouds appear to be dynamically evolving structures with short, 20-30 Myr lifetimes (see \citealt{meidtLifetimes} and references therein).  
These short lifetimes are nevertheless longer than a few free-fall times or cloud crossing times (2-10 Myr; e.g. \citealt{mckee}; \citealt{evansHeiderman}), suggesting that a simple equilibrium model can provide a satisfactory description.  

\begin{figure*}[t]
\begin{centering}
\begin{tabular}{cc}
\includegraphics[width=.485\linewidth]{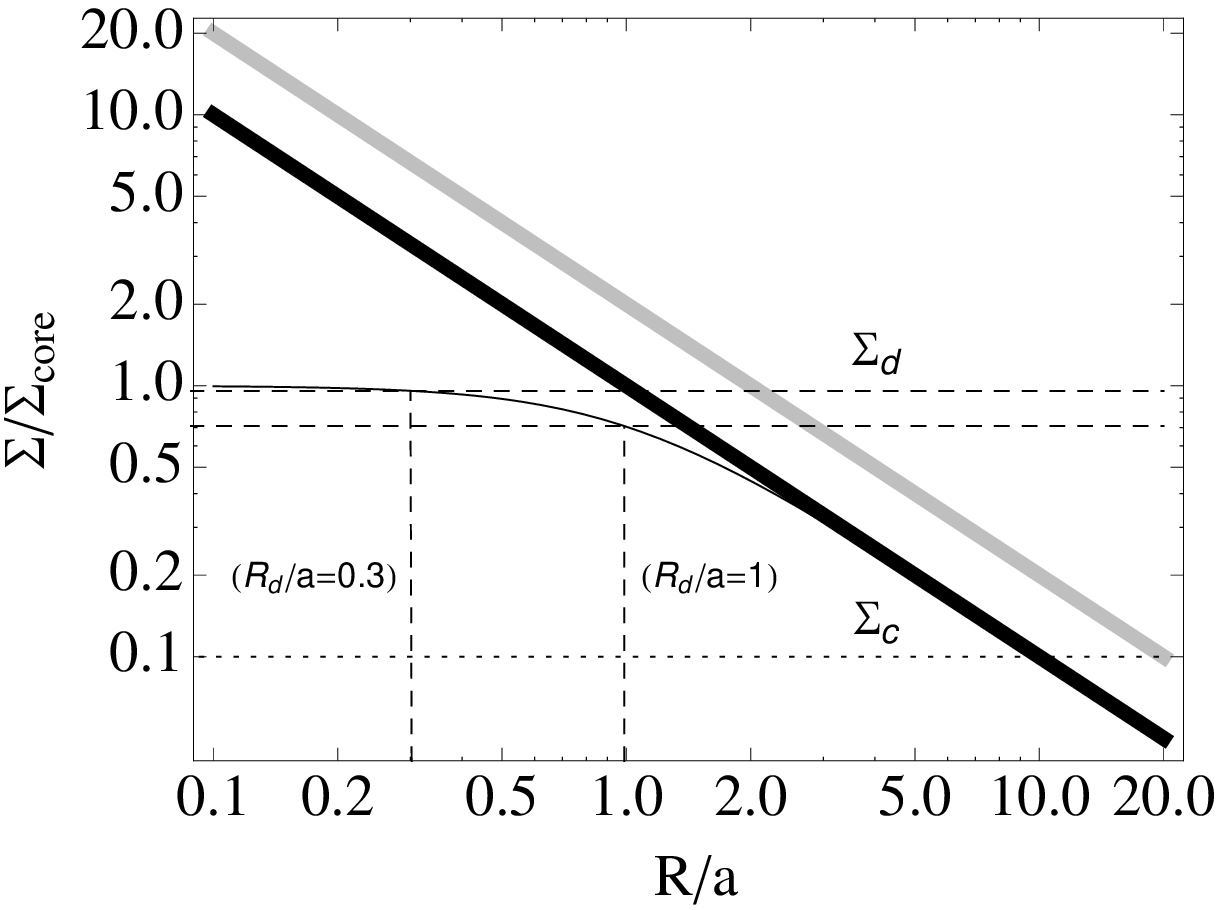}&\includegraphics[width=.485\linewidth]{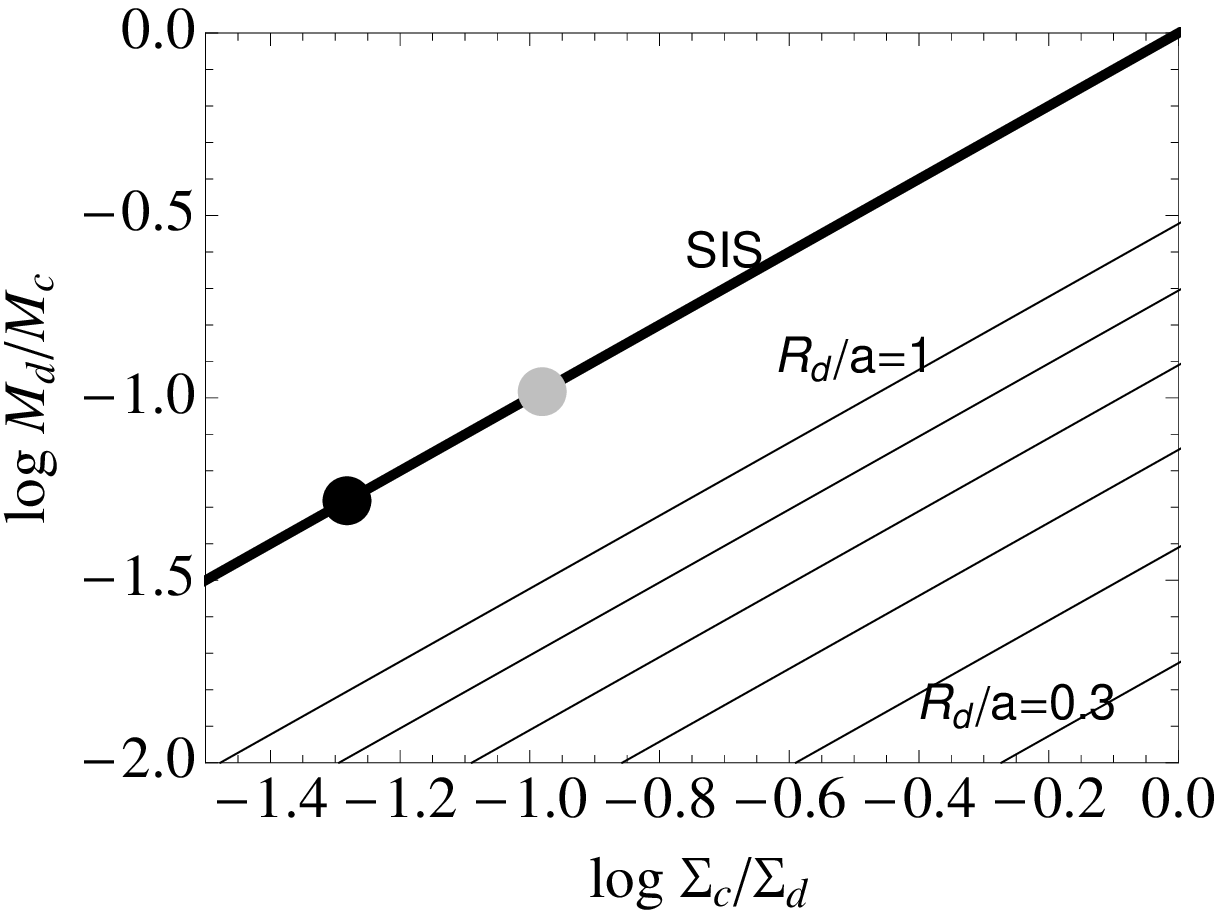}
\end{tabular}
\caption{(Left) Internal structure of the SIS (thick black and gray lines) compared to a BE sphere with core radius $a$ (thin black line).  The two SISs are shown normalized to the core radius and surface density of the example BE sphere. The intersection of each curve with the lowermost, dotted straight line and with the uppermost set of two dashed straight lines illustrate the different radial depths $R_c$ and $R_d$ reached at the given surface densities $\Sigma_c$ and $\Sigma_d$, respectively.  
Two different $\Sigma_d$ probing the BE sphere to radii $R_d$=$a$ and $R_d$=0.3 $a$ are shown. These  $\Sigma_d$ probe to varying depths in the two SISs (not indicated). (Right) Diagnostic plot of equilibrium internal structure showing the dense gas mass fraction (DGMF) $M_d/M_c$ vs $\Sigma_c/\Sigma_d$ for the cloud models on the left. The thick black solid line shows DGMFs in the SIS in black on the left measured at fixed $\Sigma_c$ as $\Sigma_d$ varies. 
(This is equivalent to the DGMF measured at fixed $\Sigma_d$ as $\Sigma_c$ varies.) The thin black lines trace the DGMFs measured in a set of BE clouds as $\Sigma_c$ varies, where each cloud is probed with a different $\Sigma_d$ (to a different depth $R_d$).  
The two filled circles illustrate the result of the different radial depths $R_d$ that can be reached at fixed surface density $\Sigma_d$ depending on the surface density of the cloud.  They correspond, in particular, to the intersection of the two SIS clouds with the larger (top-most) $\Sigma_d$ in the plot at left.  The gray point represents the more massive SIS (in gray). 
For these particular measurements, the assumed $\Sigma_c$ is the surface density of the cloud at a fixed outermost radius $R_c$=20$a$ (i.e. the gray cloud has a higher $\Sigma_c$ and global surface density). 
\label{fig:BESIS}
}
\end{centering}
\end{figure*}
\subsection{Equilibrium internal cloud structure}
\label{sec:equilStructure}
Solutions to eq. (\ref{eqn:hydro}) relate the density structure of the cloud to the difference between the external pressure $P_e$ and the pressure at any radius in the cloud $P_c$.  
This section considers both the Bonnor-Ebert (BE) sphere and the singular isothermal sphere (SIS) with density distribution $\rho\propto R^{-2}$, which is the limiting case of the BE sphere outside its central uniform density core (hereafter `BE core' to distinguish from sub-pc pre-stellar cores; \citealt{berginARAA}).  

Observations of cloud interiors are generally consistent with these equilibrium density distributions, regularly revealing a clear fall-off toward the cloud edge.  
Most typically, features like clumps and cores studied on small scales within clouds are found to agree very well with the equilibrium BE, often exhibiting a central constant density surrounded by an outer $R^{-2}$ envelope (which presumably extends to larger scales).  
But it appears much more rarely that profiles are constructed extending fully across clouds, presumably due to a combination of the geometric complexity of the cloud itself and a lack of column density tracers with sufficient dynamic range.  
Thus, internal cloud structure on scales of tens of pcs is most often assessed through measurements at two locations within the cloud, rather than with a full radial profile.  As considered in what follows, often one of these measurements probes the inner radii hosting the densest material, above a chosen 'dense gas'  surface density $\Sigma_d$$\approx$120-200 M$_{\odot} pc^{-2}$, and the second incorporates a greater fraction of the cloud material, above a characteristic 'cloud' surface density $\Sigma_c$$\approx$50-150 M$_{\odot} pc^{-2}$, and reaches further out, if not to the cloud boundary.  The two corresponding mass measurements yield a measure of the dense gas mass fraction (DGMF).

This section aims to develop a consistent basis for interpreting such measurements of internal cloud structure.  The following examines how the resemblance of observed 
clouds to one or the other case (BE sphere or SIS) depends on the depth to which a given cloud is probed.  

\subsubsection{SIS; large radius/shallow depth \label{sec:SIS}}
For many observed clouds (particularly massive ones), which tend to be probed only outside the radii at which the BE core appears, the power-law density distribution characteristic of the SIS class of solutions (the limiting case of the BE sphere) should provide a sufficiently suitable description.  

In general, at large enough radius $R$ in the BE cloud (outside the BE core), the density and mass within $R$ are given by 
\begin{equation}
\rho(R)=\frac{k^2\sigma^2}{G} R^{-2}
\end{equation}
and 
\begin{equation}
M(R)=\frac{4\pi k^2\sigma^2}{G} R
\end{equation}
where $2\pi k^2$=1 in the exact case of the SIS, in the notation adopted by \citet{dapp}.

We can therefore link the mass in the dense interior with radius $R_d$ to the mass $M_c$ inside a larger radius $R_c$: 
\begin{equation}
\frac{M_d(R_d)}{M_c(R_c)}=\frac{R_d}{R_c}=\frac{\Sigma_c(R_c)}{\Sigma_d(R_d)}  \label{eq:genratio}
\end{equation}
using the fact that $\Sigma(R)=\sigma^2/(\pi GR)$.  

As we can similarly relate the mass $M_c$ inside $R_c$ to the total mass of the cloud $M_{tot}$, the impact of the presence of external pressure becomes clear: the mass in gas at high density, above $\Sigma_d$, in the interior of the cloud is tied to the surface density $\Sigma$ of the entire cloud itself, which is set by the pressure external pressure.  

Figure \ref{fig:BESIS} shows how the mass in the dense gas $M_d$ measured in an SIS changes as a function of cloud surface density at fixed $\Sigma_d$ (the surface density above which the `dense' material sits).  On the left, a single $\Sigma_d$ can probe to varying depths in two clouds with the same self-similar density profile but different total masses within the same radius.  (Both clouds here are assumed to extend out to $R_c$=20$a$ in the figure.)  Densities above $\Sigma/\Sigma_{core}=1$ in Figure \ref{fig:BESIS} are reached at smaller radii in the cloud with lower mass and surface density  (marked in black) than in the cloud with higher mass and surface density (marked in gray).  On the right, two filled circles represent two measurements of the DGMF at fixed $\Sigma_d$, one for each of the two (black and gray) clouds. 
Note as well that a set of measurements of the dense gas fraction in a single cloud, for which $\Sigma_d$ is fixed and $\Sigma_c$ is progressively lowered (so that the $M_c$ increases), will all fall on a straight line as depicted later in Figure \ref{fig:DGMF_BH14}.

\subsubsection{BE core; intermediate depth\label{sec:BEcore}}

As clouds are probed to smaller radii, the dense gas mass may more likely reflect the presence of the BE core.  For cases in which the dense gas $R_d$ is sufficiently deep that it lies within the BE core of radius $a$, we write $M_d\approx4\pi\rho_{core} R_d^3/3$, where $\rho_{core}$ is the uniform BE core density.  
Additionally, we let $M_c$ reflect the asymptotic mass of the cored, truncated density profile adopted by \citet{dapp}, for which $k^2$=$\rho_{core}a^2G/\sigma^2$, i.e. $M_c$=$4\pi\rho_{core} a^2R_c$.  Together these yield
\begin{equation}
\frac{M_d(R_d)}{M_c(R_c)}\approx\frac{1}{3}\frac{R_d}{R_c}\left(\frac{R_d}{a}\right)^2=\frac{1}{3}\frac{\Sigma_c(R_c)}{\Sigma_d(R_d)}\left(\frac{R_d}{a}\right)^4 \label{eq:coreratio}
\end{equation}

At radii well inside the BE core, the ratio of the dense gas mass within $R_d$ to the total cloud mass within $R_c$  is more sensitive to the ratio $R_d/a$ than to $\Sigma_c$.  
As $R_d$ approaches $a$, the DGMF $M_d/M_c$ is one-third of its purely SIS value at fixed $\Sigma_c$ and $\Sigma_d$.  In the limit of $R_d$ increasing beyond $a$, though, note that we return again to the scenario described by eq.~(\ref{eq:genratio}).  

Figure \ref{fig:BESIS} illustrates the range in $M_d$ measured in a BE sphere of core radius $a$ (with mass $M_c$ and surface density $\Sigma_c$ inside a radius $R_c$) as $\Sigma_d$ varies.  
As $\Sigma_d$ increases and $R_d/a$ decreases, the linear relation shifts further below the SIS relation, to lower $M_d/M_c$.  An equivalent vertical offset (reduced $M_d/M_c$) is introduced when $\Sigma_d$ is held fixed in a set of BE clouds with varying core sizes $a$.  Clouds with broad cores will contain a lower fraction of dense gas than their narrow-cored counterparts, which approach the SIS.

\subsubsection{Additional (non-equilibrium) structure\label{sec:noneqstruct}}
The structure of clouds at the highest densities, where clumps and cores are observed, is most certainly more complex than parameterized by either the idealized SIS or BE sphere. \\  
\indent{\it point-like masses; innermost radii}\\
The very interiors of clouds may be characterized by structures that arise with departures from equilibrium (collapse) on the smallest scales.  These features may be present in all clouds, but their contribution to $M_d$ can become non-negligible when $\Sigma_d$ probes the inner depths, such as in clouds with especially narrow BE cores.   
At such small radii, observations are more likely to reveal these features in the cloud structure.  In these cases, the dense gas mass will exceed that expected for the SIS as explored more in the next section.      

{\it uniform/shallow density profiles}\\
Some clouds may also resemble uniform density structures, either because they are genuinely out of hydrostatic equilibrium or because measurements are confined to within or very near the BE core.
When $\Sigma_c$ itself approaches the BE core (or, conversely, when the core is very broad), the dense gas mass fraction behaves nearly independently of $\Sigma_c$, as it would in a uniform-density cloud.  

The shallow $r^{-3/2}$ profile characteristic of Lin-Shu collapse (see $\S$ \ref{sec:collapse}) from a state of hydrostatic equilibrium likewise leads to rapid changes in $M_d/M_c$.  
In general, at any radius from within a shallow density distribution $\rho\propto R^{-n}$ with $n$$>$1 \begin{equation}
\frac{M_d(R_d)}{M_c(R_c)}\propto\left(\frac{\Sigma_c(R_c)}{\Sigma_d(R_d)}\right)^\frac{3-n}{n-1}
\end{equation}
Thus inside the $r^{-3/2}$ core of a collapsed cloud $M_d(R_d)/M_c(R_c)\propto\Sigma_c^3$.  Note, though, that 
beyond the collapse radius of such a cloud the profile is expected to be described by an $R^{-2}$ tail.  Therefore when $R_c$ probes out in the tail while $R_d$ falls within the $r^{-3/2}$ collapsed core of radius $a$ then
\begin{equation}
\frac{M_d(R_d)}{M_c(R_c)}\propto\frac{2}{3}\left(\frac{R_d}{a}\right)^2\left(\frac{\Sigma_c(R_c)}{\Sigma_d(R_d)}\right)\label{eq:hybriddens}
\end{equation}
using that $R_c=\Sigma_d/\Sigma_c (R_d a)^{1/2}$ as can be derived by requiring that the inner and outer densities match at radius $a$. 

Note that the dependence on $\Sigma_c$ in eq. (\ref{eq:hybriddens}) is almost indistinguishable from that of a Bonnor-Ebert sphere.  
$\S$ \ref{sec:collapse} later discusses distinguishing between equilibrium and collapse.  
\subsection{Relation to average cloud surface density}
Given the inexact nature of the cloud boundary inferred through observation, the direct connection between the dense mass and total mass is more challenging than linking the dense mass to the mass inside an arbitrary boundary, defined either as an empirically-chosen threshold $\Sigma_c$ or $R_c$, as written in eq. \ref{eq:genratio}.  More typically, the cloud surface density is estimated from the measured cloud mass and size, 
i.e  
\begin{equation}
\overline{\Sigma}_{c}=\frac{M_c}{\pi R_c^2}, \label{eq:surfdens}
\end{equation}
using techniques that decompose the emission from a particular molecular tracer in position-position-velocity space (i.e. CPROPS; \citealt{CPROPS}) into individual cloud entities.  
Since the mean surface density inside $R_c$ is $\overline{\Sigma}_{c}$=$2\Sigma_c(R_c)$ for the SIS cloud (and in the limit of large $R$ generally for the BE sphere), we write eq. (\ref{eq:genratio}) as 
\begin{equation}
\frac{M_d}{M_c}=\frac{\overline{\Sigma}_{c}}{2\Sigma_d} \label{eq:finratio}
\end{equation}
with the radial dependence of all quantities implicit.  

To evaluate eq. (\ref{eq:finratio}), the only requirement is that the cloud mass $M_c$ must relate to the surface density $\overline{\Sigma}_{c}$ self-consistently; neither is required to represent a total quantity, even if the manner in which they are measured is for this purpose (i.e. via extrapolation below the initial detection threshold).  

The dense gas mass $M_d$ in eq. (\ref{eq:finratio}) must represent all mass above the threshold surface density $\Sigma_d$, such as typically inferred from observations of dust in extinction or emission (e.g.  \citealt{heiderman}; \citealt{battistiHeyer}).  When the dense gas is traced via molecular line emission so that the measured dense gas mass and surface density are related as expressed by eq. (\ref{eq:surfdens}), then eq. (\ref{eq:finratio}) loses the factor of 2 and $M_d/M_c$=$\overline{\Sigma}_{c}/\overline{\Sigma}_d$  in a SIS.  

\subsection{Predicted trends within cloud populations}
According to the previous section, populations of clouds with a spread in mass and surface density will exhibit a range of DGMFs simply due to the varying depths to which the clouds are probed, which impacts the resemblance to one or the other case described there.  
Figure \ref{fig:BESIS} already showed for the example of a SIS that a single $\Sigma_d$ can probe to varying depths in clouds with different $\Sigma_c$.  

\begin{figure*}[t]
\begin{centering}
\begin{tabular}{c}
\includegraphics[width=.65\linewidth]{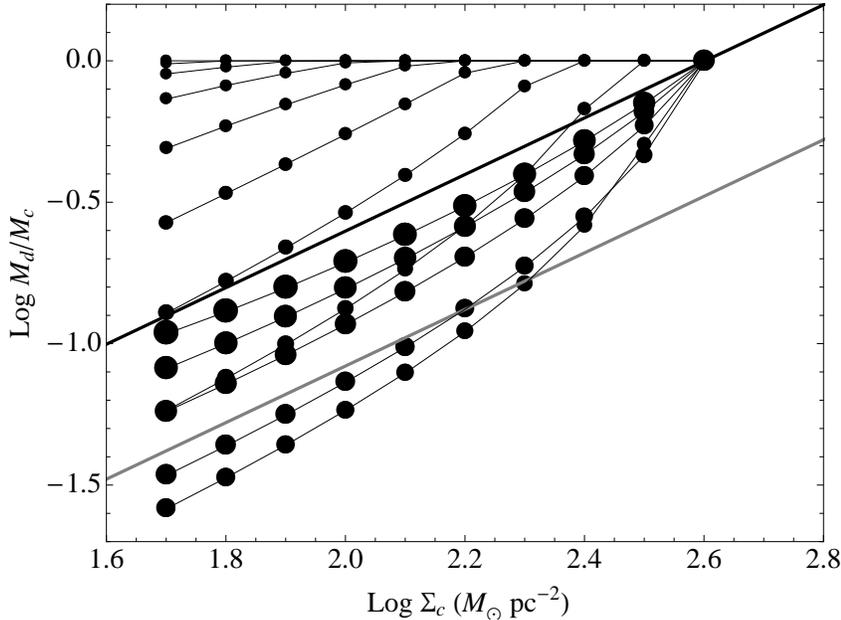}\end{tabular}

\caption{Plot of $\log{(M_d/M_c)}$ vs. $\log{\Sigma_{c}}$ 
for a model cloud population with varying surface density (and total mass and size), in which the dense gas mass is measured above a fixed $\Sigma_d$=200 M$_{\odot} pc^{-2}$.  The size of the point increases with the cloud mass $M_c$. 
The black line depicts the model relation $M_d/M_c$=$(\Sigma_{c}/\Sigma_{d})$ corresponding to the asymptotic SIS (far outside an interior BE core), while the gray line shows the expected relation when $\Sigma_d$ lies just at the BE core edge, $M_d/M_c$=$(\Sigma_{c}/3\Sigma_{d})$. Both cases adopt $\Sigma_d$=200 M$_{\odot} pc^{-2}$.  
\label{fig:DGMF_model}
}
\end{centering}
\end{figure*}

Figure \ref{fig:DGMF_model} explores how the features in a more general (non-SIS) model for internal cloud structure lead to variations in the dense gas mass fraction when $\Sigma_d$ and $\Sigma_c$ each probe to varying depth.  We consider a population of BE-like clouds in which the shape of the density profile is held fixed but scales up or down according to the cloud's total mass.  
Thus the BE core radius, which ranges  from 1-5 pc in the population, increases together with cloud mass, which spans 10$^{3.5}$ to 10$^{5.5}$ $M_\odot$.  In this model, clouds with different masses can have the same $\Sigma_c$, depending on the 'pronouncedness' of the BE core in their density profile.  
Additionally, each cloud contains a central uniform sphere of mass 10 $M_\odot$ inside a fixed radius $a_p$=0.1 pc in the flattened BE core, which represents a collapsed (prestellar) core.  

Figure \ref{fig:DGMF_model} exhibits several features which should be characteristic of cloud populations in general.   
First, in the most massive clouds both  $\Sigma_d$ and $\Sigma_c$ probe the SIS power-law tail, but as the cloud mass decreases, $\Sigma_d$ begins to fall within the core.  In this case clouds are in the `intermediate depth' regime and fall below the SIS line.  
Second, when $\Sigma_c$ is high enough, a subset of clouds are in the `uniform density' regime and $R_c$ itself falls near or within the core, i.e $\Sigma_c$$\approx$$\Sigma_{core}$.  
At fixed mass, the relation between $M_d/M_c$ and $\Sigma_c$ thus steepens, as described in $\S$ \ref{sec:noneqstruct}.    
Third, especially in low mass clouds (here, those with $M$$\lesssim$10$^{4} M_{\odot}$), $\Sigma_d$ can reach deep enough that $M_d$ begins to reflect the presence of the central point mass, thus raising the dense gas mass fraction above the level expected for the equilibrium SIS.  

Another notable feature of Figure \ref{fig:DGMF_model} is the characteristic deficit of clouds with high surface density and low dense gas mass fraction, giving the impression of an upward trend. Thus, overall, the cloud population can be well-described simply by the SIS, with scatter that reflects real departures from the idealized, equilibrium density profile.  

In similar populations, clouds with relatively low $M_d/M_c$ at fixed $\Sigma_c$ may indicate that the persistence of the power-law in to the region of high density may not be the most realistic, i.e. if the density profile has already developed a core at the threshold density $\Sigma_d$. 
Clouds with relatively high $M_d/M_c$ that fill in the region of parameter space sitting above the SIS relation (i.e. with low masses and low surface densities) meanwhile likely indicate that central structures not associated with equilibrium are being probed.  
Extragalactic cloud surveys will contain fewer of these low mass clouds than observable in the MW (with masses $M$$\lesssim$10$^4 M_{\odot}$, given the surface brightness limits and resolutions achieved by modern facilities; e.g. \citealt{bolatto}, \citealt{hughesII}). Thus a much narrower spread around the SIS prediction should be expected to be observed in nearby galaxies. 

\subsection{Distinction from collapse}
\label{sec:collapse}
The pressurized cloud model relates the dense gas mass contained within clouds to the internal density structure that emerges specifically under hydrostatic equilibrium. Observations in agreement with the model therefore arguably supply evidence for the role of external pressure.  
But as noted previously, the increasing densities that emerge toward the center of a collapsing spherical cloud resemble the hydrostatic equilibrium profile, making these two scenarios difficult to distinguish based on morphology alone.  

The collapse process has been studied in a variety of ways, both with analytical similarity solutions and numerically, and starting from a variety of initial and boundary conditions (e.g., \citealt{larson}; \citealt{penston1}; \citealt{penston2}; \citealt{shu77}; \citealt{foster}; \citealt{hennebelle}).  The range of possible outcomes are thought to be bounded by two limiting scenarios \citep{hunter}.  
Larson-Penston collapse begins at initially uniform density and develops a BE-like structure with the same asymptotic scaling as the SIS, but with 4.4 times larger density.  
As in the case of the free-fall collapse of a unifom density sphere, the outer regions fall in faster than the inner regions so that all infalling shells reach the center at the same time. 

The "inside-out" collapse solution, which was developed by \citet{shu77} to have more realistic initial conditions, begins in the equilibrium state described by the SIS and generates an $r^{-3/2}$ density profile followed by an $r^{-2}$ tail.  
Like Larson-Penston collapse, the inside-out solution is virtually indistinguishable from hydrostatic equilibrium without additional kinematic information.  But unlike Larson-Penston collapse it is arguably still consistent with the pressurized cloud model, as clouds begin in equilibrium.  

Larson-Penston collapse, on the other hand, seems unlikely simply due to the large infall velocities (-3.3 times the isothermal sound speed) at large radii.  As a result of such fast infall, the density, velocity, and  mass accretion rate are larger in the LP solution than in the Shu solution at similar locations within the cloud. 
While collapse of this kind may be descriptive on the scales of clumps and cores (but cf. \citealt{ketoCR}), it seems inconsistent with observations on cloud scales.
Indeed, the collapse of entire clouds would seem inconsistent with the relative inefficiency of star formation, as first argued by \citealt{ZuckermanEvans}; if all clouds are presently undergoing collapse, observed star formation rates in the MW should be much higher.  

Thus observed supersonic line widths in molecular clouds are interpreted here as arising with turbulent motions that support clouds, rather than to large-scale infall \citep{ZuckermanEvans}, and thus to the idea that clouds are in rough hydrostatic equilibrium.   
The model does allow for collapse on small scales (see $\S$ \ref{sec:noneqstruct}), i.e. within the densest cloud interiors, such as may be characteristic of the formation of clumps and cores. But this is expected to be characteristic at radii typically well inside those probed by the threshold $\Sigma_d$.   

\subsubsection{Relation to the turbulent picture}
The model developed in this paper is not unlike most studies of internal cloud structure and collapse in which it is common to assume static initial configurations with either uniform density or BE hydrostatic equilibrium profiles (e.g. as reviewed by \citealt{gammie}, \citealt{BP}, \citealt{ketoCR}).  
But as molecular clouds and the cores within them are thought to be supersonically turbulent (given their supersonic linewidths; \citealt{ZuckermanPalmer}), it has been suggested that they are the product of turbulent density fluctuations or supersonic compressions within their environments (e.g. \citealt{ballesteros}).  
In such a scenario it seems unclear whether hydrostatic equilibrium is applicable or how, in such turbulent conditions, such clouds might develop.  

Arguably, however, the build-up of high densities and the development of structure within the ISM as described by turbulence is not necessarily exclusive of the concept of local pressure equilibrium.  
Clouds created by turbulent compression that are at least initially confined by the ambient non-thermal pressure in the ISM may very well resemble marginally stable BE spheres or SISs (with the boundary pressure set by the ambient turbulent pressure; \citealt{larson03}).  
(The same would apply at smaller scales, to star-forming cores within clouds, which would be pressurized by the ambient non-thermal pressure within the cloud itself.) 

The idea proposed here that environment determines the cloud dense gas mass fraction is thus complimentary to the studies of \citet{kain2009}, \citet{kaintan} and \citet{kain2013}, in which turbulent driving is thought to be responsible for the accumulation of mass at high densities in clouds through compression.  
In both cases, the exterior of the cloud is critical for establishing internal cloud structure.  Our picture would apply a pressure-based description to the structure developed at the compression.  

The relevance of a micro-turbulent framework could be tested by probing the role of external pressure in setting the cloud gas surface density, as suggested as well in $\S$ \ref{sec:Pdiscussion}.  Again, though, it should be noted that the clouds in M51 indirectly support such a picture, in that clouds appear to inherit their properties from their environment (\citealt{colombo2014a}; \citealt{hughesII}).  

\begin{figure*}[t]
\begin{centering}
\begin{tabular}{c}
\includegraphics[width=.75\linewidth]{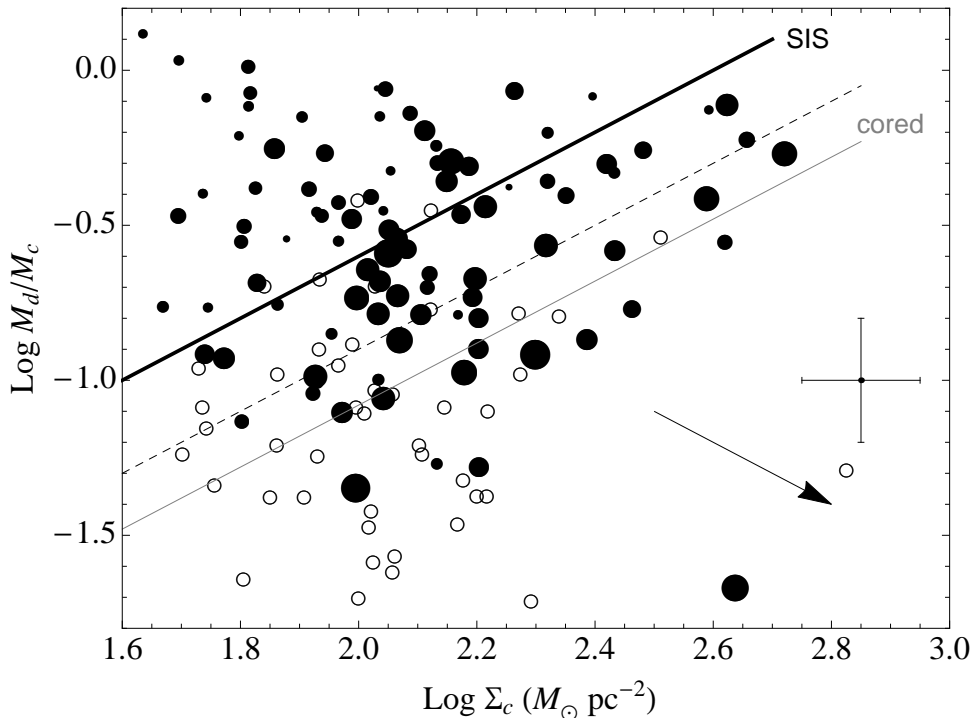}
\end{tabular}
\caption{Plot of $\log{(M_d/M_c)}$ vs. $\log{\Sigma_{c}}$ 
for the subsample of 141 clouds selected from the \bhalias~ sample, with S/N$>$2 and containing a single BGPS source.  Measurements of $M_d$ for all of these clouds are made above a fixed $\Sigma_d$=200 M$_{\odot} pc^{-2}$.  The sizes of the solid black points increase with the 
cloud mass $M_c$ in the range 10$^{3.2}$$<$$M_c/M_{\odot}$$<$10$^{5.8}$.  
Open circles mark clouds with large angular sizes $>$490", in which there is the greatest potential for underestimation in the measured $M_d$ due to median-filtering of the 1.1 mm dust emission tracing the dense gas.  
A representative error bar reflecting the measurement uncertainties tabulated by BH14 is shown in the bottom right corner.  
The black line depicts the model relation $M_d/M_c$=$(\Sigma_{c}/\Sigma_{d})$ corresponding to the asymptotic SIS (far outside an interior BE core), while the gray line shows the expected relation when $\Sigma_d$ lies just at the BE core edge, $M_d/M_c$=$(\Sigma_{c}/3\Sigma_{d})$. 
Both cases adopt $\Sigma_d$=200 M$_{\odot} pc^{-2}$.  The dashed line shows the best-fit relation $M_d/M_c$=$(\Sigma_{c}/\Sigma_{d})$ with $\Sigma_d$ free.  The black arrow in the bottom right corner represents a factor of two increase in cloud mass.  \label{fig:DGMF_BH14}
}
\end{centering}
\end{figure*}
\section{Comparison to observed internal cloud structure}
\label{sec:tests}
According to the previous section, the pressure exerted from exterior to the clouds sets up a pressure gradient within the cloud, thus introducing a strong density contrast from the inside to the edge\footnote{Here `edge' refers to the radius at which the cloud's internal and external pressures are equal.} of the cloud.  
This section examines the clouds studied by Battisti \& Heyer (2014; hereafter \bhalias) for consistency with this simplified model.  

\subsection{Notes on the BH14 cloud sample}
BH14 use a unique combination of tracers and techniques to measure the dense gas fraction in a sample of over 400 clouds in the Bolocam Galactic Plane Survey (BGPS; \citealt{aguirre}). They estimate the bulk mass of each cloud following its identification in $^{13}$CO J = 1 - 0  emission with the CPROPS \citep{CPROPS} cloud decomposition algorithm (see BH14 for details). This mass, together with the extrapolated cloud size, defines the surface density as expressed here in eq. \ref{eq:surfdens}. 
The dense gas mass estimated by BH14 and used here is defined by the mass in confined\footnote{
The dust sources, which are required to coincide with spectroscopic detections of surveyed dense gas tracers, are matched to CO emitting objects along the line-of-sight kinematically (see BH14 for details).  To assign the kinematic distance of each CO emitting object, and disambiguate between its location on the near or far side of the galaxy, BH14 use a combination of HI absorption along the line of sight and expectations for a well-defined size-linewidth relation amongst clouds.   
} 
FIR BGPS sources above a fixed 1.1mm surface brightness.   
This corresponds to a uniform column density threshold assuming a fixed dust-to-gas ratio, dust temperature and emissivity.  This threshold is what is called $\Sigma_d$ above and is therefore the same for all clouds in the sample.  
BH14 estimate this threshold to be $\Sigma_d$=200 M$_\odot pc^{-2}$.  According to BH14, dust sources tracing cloud cores and clumps are marginally to well-resolved by the 33" Bolocam resolution, which corresponds to a physical scale between 0.3 and 2 pc at the distances of the sources (see Fig. \ref{fig:angsizeDplot}).  

The combination of techniques used by BH14 is key to the size and spatial coverage of the sample, which 
spans a larger dynamic range in cloud surface density and probes a wider range in environment (external pressure) than typical of smaller samples directed within the Solar neighborhood.  
The use of CO as cloud tracer allows a more complete inventory than possible, e.g., tracing clouds via dust extinction (which is limited to only on relatively nearby clouds).  
But a comparative disadvantage is that the dense gas mass fraction for a given cloud is based on heterogeneous mass estimates, whereas dust extinction can probe column densities and masses from throughout a cloud.  
Additional inhomegeneity from cloud to cloud can be introduced given the uncertainties associated with converting a uniform FIR flux into a uniform gas surface density $\Sigma_d$, 
which can be sensitive to the assumed dust temperature (and emissivity).  

According to BH14 (and as discussed further in the Appendix), we expect systematic uncertainties on the dense gas mass to be no larger than a factor of 2 (BH14).  (BH14 estimate random measurement errors on the order of 20\%; see representative error bar here in Figure \ref{fig:DGMF_BH14}.)  The `bulk' cloud masses and surface densities from CO are similarly good to within a factor of 2.  These uncertainties should not obscure trends exhibited by the ensemble of BH14 clouds, given the large (one order of magnitude) dynamic range in $\Sigma_c$ within the sample. 

Like BH14, only clouds for which the cloud mass $M_c$ exceeds twice the random measurement uncertainty (S/N$>$2) are considered here.  The present study is also limited to only those clouds that contain a single BGPS source.  
Multiple FIR sources interior to a single CO-emitting object (or possibly multiple sources along the line of sight) represents a more complex, multiple cloud structure than the nominal configuration elected for testing here, namely a centralized region of high density inside a single cloud.  
Perhaps more relevantly, the 68\% of the sample that contains multiple sources are consistently objects with the largest angular sizes and masses, as discussed in the Appendix (see Figure \ref{fig:angsizeDplot}).  
As noted by BH14, median filtering to subtract the FIR background at 1.1 mm can preferentially remove real extended low-level emission from the dust in such large sources.  \citet{aguirre} estimate that flux recovery can be quite low with this procedure, less than 20\% for sources with sizes $\theta_{FWHM}$$>$350".  
(Indeed, dense gas mass fractions in these large objects with multiple FIR sources are found to be consistently lower (by a factor of $\sim$3) than in smaller clouds.)  
By removing these objects from the sample, we should be left with clouds in which the filtering has less impact and at least more uniformly affects the inferred dust mass distribution from cloud to cloud.  
The impact of median filtering on the dense gas mass fractions measured in the remaining clouds is discussed in the Appendix.  

\subsection{Trends in the dense gas mass fraction}
\label{sec:DGMFtrends}
Figure \ref{fig:DGMF_BH14} shows the dense gas fraction plotted against the cloud surface density for the sub-sample of 141 clouds selected in the previous section on a log-log scale.  The ensemble of clouds tends to trace an increase in $M_d/M_c$ with increasing $\Sigma_c$, though with some scatter, especially at low cloud surface densities.  
At high surface densities, Figure \ref{fig:DGMF_BH14} clearly displays a notable lack of clouds with low dense gas mass fractions.  This region of parameter space is under-populated even considering the entire BH14 sample of clouds.  

According to eq. (\ref{eq:finratio}), clouds resembling the SIS should trace out a linear relation in Figure \ref{fig:DGMF_BH14} with a well-defined intercept of -$\log{2\Sigma_d}$, given that the threshold $\Sigma_d$ is expected to be the same for all clouds.  The best-fit linear relation to the points in Figure \ref{fig:DGMF_BH14} suggest $\Sigma_d$=400 M$_\odot$pc$^{-2}$, only a factor of 2 from the value $\Sigma_d$=200 M$_\odot$pc$^{-2}$ we can expect to retrieve, which could itself admit a factor of 2 uncertainty (BH14).  

On the other hand, eq. (\ref{eq:finratio}) is expected to apply only as long as clouds are sampled well outside any BE core that may exist.  
Genuine departures from a single linear relation are expected to arise given deviation from the self-similar SIS power-law regime, as described in the previous section.  
Thus, the relatively low fitted intercept could also imply that, throughout much of the sample, $\Sigma_d$ probes sufficiently deep within the cloud that the dense gas mass lies primarily within the constant density core.  (This would reduce the DGMF relative to the SIS case.)  
In the event that the $\Sigma_d$ threshold occurs exactly at the outer edge of the BE core (so that $R_d$=$a$), then eq. (\ref{eq:coreratio}) with $\overline{\Sigma}_c$=2$\Sigma$ implies that the intercept would be reduced by 0.48 dex from the asymptotic SIS case, to -$\log{6\Sigma_d}$.  Eq. (\ref{eq:coreratio}) thus might also imply that $R_d$$\sim$1.1$a$ on average in the sample, based on the best-fit intercept.  

As outlined in the previous section and parameterized in the model, ambiguity in the actual shape of the density profile, and how far into (or beyond) the BE core the dense gas probes produces scatter in Figure \ref{fig:DGMF_BH14}.  
This can explain the presence of objects in the region of parameter space that lies above the SIS line in Fig. \ref{fig:DGMF_BH14} even with a stricter S/N criterion. According to our simple model, clouds that lie significantly above the SIS line in Figures \ref{fig:DGMF_model} and \ref{fig:DGMF_BH14}, particularly at low $\Sigma_c$, can be expected when, as here, they are preferentially low-mass.  In this case, $\Sigma_d$ can reach very deep within the cloud.  
When $\Sigma_d$ probes very small radii, the dense gas mass is more likely dominated by small-scale structure that could be related to collapsed (or collapsing) objects, i.e. clumps and cores, or to fragmentation and the growth of instabilities in the constant density core.  Such structure could be present in all clouds, but not contribute significantly in high mass objects. 

The above explanation for the enhanced DGMFs observed is by no means unique.  
But attributing them to the global collapse of clouds, which could, e.g., lead to continued build-up of centrally high densities, seems less compelling here: there is no immediately obvious reason why global collapse should be favored in low mass clouds any more than in their high-mass counterparts, as would be required to explain Figure \ref{fig:DGMF_BH14}. 
This would favor the interpretation suggested in $\S$ \ref{sec:equilStructure}, in which DGMFs that are enhanced (relative to the SIS case) are not the result of the global collapse of the cloud but rather stem from local collapse, on small, core/clump scales.  

Overall, the agreement between the internal structure of BH14 clouds and the model prediction in Figure \ref{fig:DGMF_model} presents compelling evidence that clouds are in rough hydrostatic equilibrium in the presence of an external pressure.   
The consistency with the SIS is quite strong (a linear relation between dense gas fraction and $\Sigma_c$ is not ruled out by the data) and the lack of clouds with low $M_d$ in the bottom right corner of the plot, in particular, would seem to strongly imply that the internal structure of clouds is being regulated in the manner proposed in $\S$ \ref{sec:model}.  
\subsubsection{Other sources of scatter}
By comparison with Figure \ref{fig:DGMF_model}, we can attribute most of the scatter about the predicted linear relation between $M_d/M_c$ and $\Sigma_c$ to variation in the dominant shape of the density profile within the $\Sigma_d$ threshold, from cored (BE) to strictly power-law (SIS).  
But there are other sources of scatter, part of which can be associated with observational uncertainties and part that may track real departures from our simple model, as discussed in the Appendix. 

It should be emphasized, however, that observational uncertainties are not driving the trend (even if they may contribute to the scatter).  Uncertainties on $M_d$ and $M_c$ for this sample are estimated to be at maximum a factor of 2, or about 0.3 dex, whereas the linear trend can be traced over a full 1 dex in $\Sigma_c$ and almost 1.5 dex in $M_d/M_c$.  
Note as well that the shared dependence of $\Sigma_c$ and $M_d/M_c$ on $M_c$, which could lead to correlated uncertainties, would introduce a trend in the opposite sense of the overall trend exhibited in Figure \ref{fig:DGMF_BH14}, as shown by the black vector.   

In addition, DGMFs in the clouds that sit above the SIS lie (which appear to contain almost as much dense gas as the mass in the rest of the cloud), may be lower, e.g. if the cloud edge extends beyond the $R_c$ measured from $^{13}$CO emission.  
Again, this does not invalidate eq. (\ref{eq:finratio}), which applies among clouds of various sizes (and surface densities) as long as $\Sigma_d$ is fixed, as well as within a given cloud, at different $\Sigma_c$ (see Section \ref{sec:equilStructure}). 

\subsubsection{Relation to other work}
Battisti \& Heyer (2014) argue from these data that there is no dependence of dense gas mass fraction on cloud mass or surface density.  
However, this was based on the consideration of all 438 catalogued clouds with matched BGPS sources, including those with multiple sources and large angular areas that significantly increase the scatter in the parameter space (see Appendix).  
For many of these, the DGMF is likely significantly underestimated.  As a result, not only is the trend with $\Sigma_c$ found here (irregardless of interpretation) missed, but the sub-population of clouds with very high dense gas fractions (only 53 out of 438 above the SIS line) tend to be de-emphasized.  

The present study therefore does not strongly support the conclusion of Battisti \& Heyer (2014) that the low dense gas fractions exhibited by MW clouds explains the low observed star formation efficiency in the MW and other galaxies.  
In this picture, clouds convert only a small fraction of their mass in to stars per free-fall time.  But very clearly, some clouds have very high dense gas fractions, implying high conversion efficiency.  
Instead, the alternative explanation for low global SFEs seems favorable, in which some clouds in a given population are not actively star-forming, while others can be as highly efficient as often found in simulations (e.g. \citealt{bonnell03}; \citealt{klessenB}; \citealt{VS}).  
Such a picture is supported by galaxy-scale models of GMC formation and evolution (i.e.  \citealt{dobbsPringle}) and measurements of short GMC lifetimes in nearby galaxies (e.g. \citealt{blitz}; \citealt{kawamura}; \citealt{meidtLifetimes}), which imply that clouds are dynamically evolving structures. 
\begin{figure*}[t]
\begin{centering}
\begin{tabular}{ccc}
\includegraphics[width=.485\linewidth]{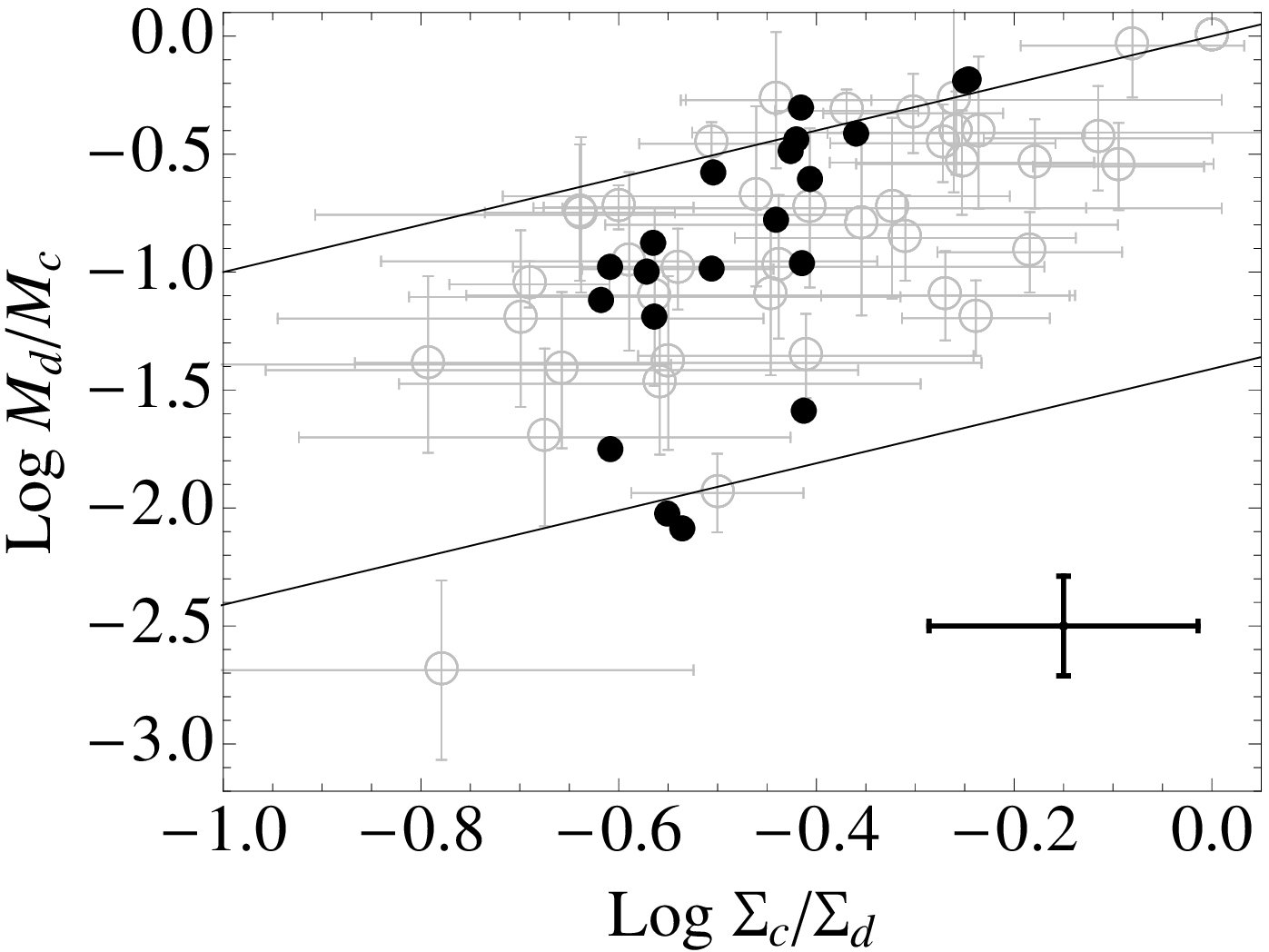}&
\includegraphics[width=.485\linewidth]{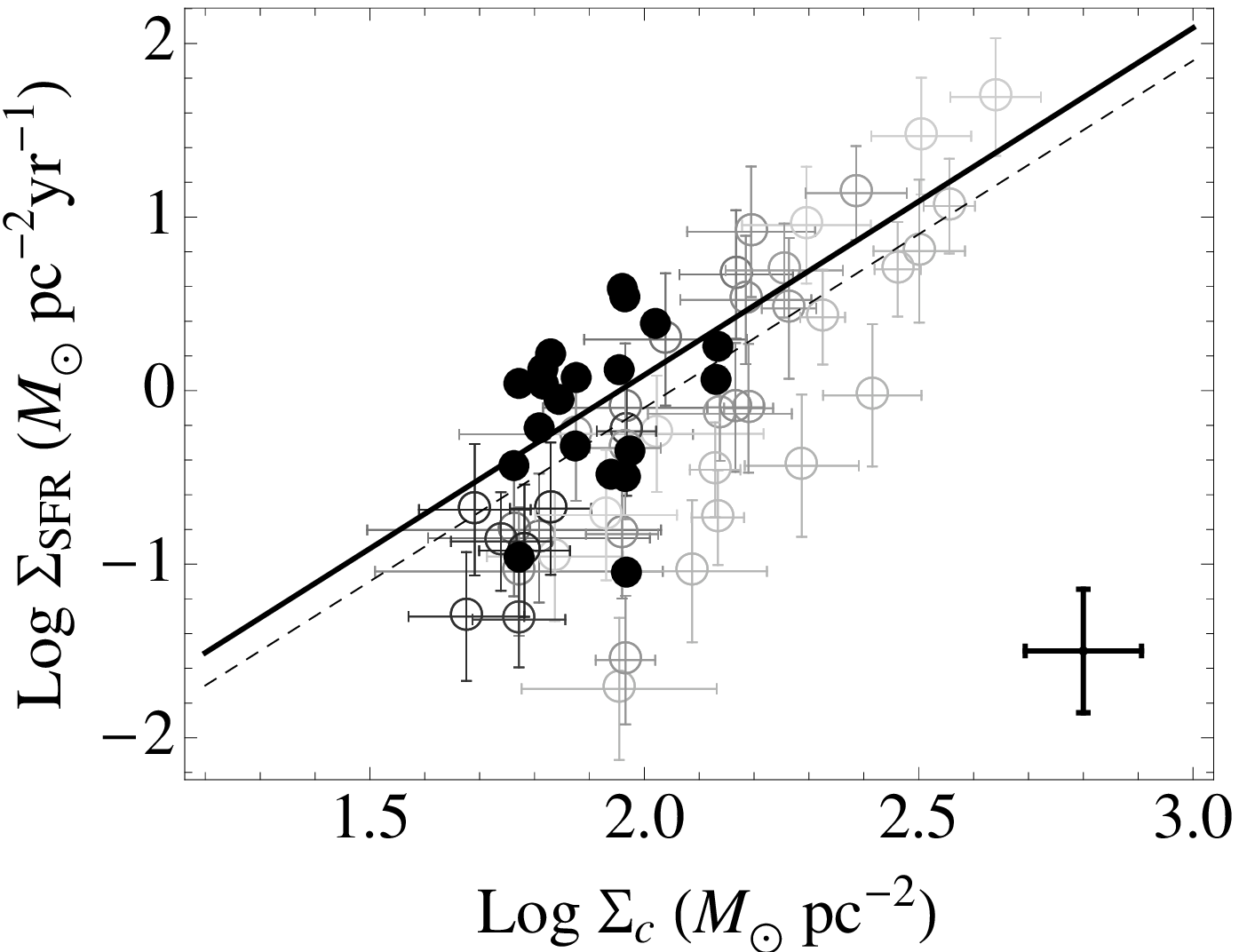}
\end{tabular}
\caption{(Left) Plot of $\log{(M_d/M_c)}$ vs. $\log{\Sigma_{c}/\Sigma_{d}}$ 
within 20 Solar neighborhood clouds studied by \citet{heiderman} and \citet{evansHeiderman}.  The black solid points show dense gas measurements from \citet{evansHeiderman} for all 20 clouds extracted above a fixed dust extinction threshold $A_v$=8 corresponding to $\Sigma_d$=120 M$_\odot$ pc$^{-2}$ (see text).  
Open symbols show measurements extracted from {\it within} the clouds, at multiple $\Sigma_d$ defined by contours of dust extinction typically ranging from $A_v$=8 to $A_v$=20 \citep{heiderman}.  A single value for $\Sigma_c$ is taken from the lowest tabulated contour level in each cloud.  
Between 3 and 6 contour levels are defined per cloud (as tabulated by \citet{heiderman}al), thus a single cloud contributes multiple data points in the plot.  
A representative error bar for the black symbols (defined by taking the average of the uncertainties in the measurements from throughout clouds reported by \citealt{heiderman}) is shown in the bottom right corner.  
The two lines show the model relation for an SIS (top) and the prediction for a BE sphere  probed to depth $R=0.6 a$ repeated from Figure \ref{fig:DGMF_model} (bottom).  
(Right) Plot of $\log{(\Sigma_{SFR})}$ vs. $\Sigma_{c}$ for the same 20 nearby clouds shown at left.  Black closed symbols show measurements from all clouds made with a fixed $\Sigma_d$=120 M$_\odot$ pc$^{-2}$.  
A representative error bar for the black symbols (defined by taking the average of the uncertainties in the measurements from throughout clouds reported by \citealt{heiderman}) is shown in the bottom right corner. 
Open symbols show measurements probing throughout the clouds, letting each extinction contour reported by \citet{heiderman} supply a measure of $\Sigma_c$ and using the highest probed contour level to define a single $\Sigma_{d}$, SFR and dense gas mass per cloud.  
$\Sigma_d$ varies from cloud to cloud, as represented by the gray-scale of the points, which ranges from $\Sigma_d\approx$100 (black) to 500 (light gray).   
The two lines show the best-fit relation $\Sigma_{SFR}$=$\epsilon\Sigma_{c}^2/\Sigma_{d}$, with $\Sigma_d$ free, determined by fitting only to the solid symbols (solid black) or only to the open symbols (dashed).  
\label{fig:heidermanSFR}
}
\end{centering}
\end{figure*}

\section{Implications for star formation relations}
 \label{sec:SFimplications}
\subsection{Cloud-scale star formation}
\label{sec:localSF}
In the pressurized cloud model, the cloud's surroundings establishes its internal structure and the dense gas fraction is set by the cloud surface density.  
According to several observational studies, the rate at which a cloud forms stars is most closely related to the mass of dense gas within it (\citealt{johnstone}; \citealt{andre}; \citealt{li97}; \citealt{lada92}; \citealt{wu05}; \citealt{heiderman}; \bhalias).  Thus, in this light, the cloud environment 
is arguably the basis that determines the rate at which a cloud forms stars.  

This framework has direct implications for the observed relation between gas and star formation.  Specifically, if the mass in dense gas drives star formation, then there should be a connection between star formation and cloud surface density. 
Allowing that the SFR $\dot{M_\star}$=$\epsilon M_d$ with efficiency $\epsilon$ and, from eq. (\ref{eq:finratio}), 
 \begin{equation}
 M_d=\frac{M_c \Sigma_c}{\Sigma_d} 
 \end{equation}
then the star formation rate per unit cloud area $\dot{M_\star}/\pi R_c^2$ (the star formation rate surface density)
\begin{equation}
 \Sigma_{SFR}=\epsilon\frac{\overline{\Sigma}_c^2}{2\Sigma_d} \label{eq:sfrelation}
\end{equation}
where $\Sigma_d$ here is the surface density above which the (dense) cloud material is star forming.    
Here, as in the \citet{ostrikerShetty} picture, pressure equilibrium-regulated star formation leads to a  quadratic dependence of the star formation rate on the gas surface density.   

Note that, as thoroughly discussed in $\S$ \ref{sec:equilStructure}, real clouds in equilibrium may show departures from this simplified relation, i.e. when observations reveal more complex internal structure than parameterized by the SIS as assumed.  This should be most common for the dense star-forming material in low mass clouds when, as inferred for such clouds in the BH14 cloud sample, the dense gas threshold $\Sigma_d$ probes far in to the BE core.  
Eq. (\ref{eq:sfrelation}) should be largely applicable at higher cloud masses, and in the extragalactic context, where observations are dominated by more massive, high surface density clouds in the population.  

In general, as long as the properties of the star-forming material (mass, star formation rate) within all clouds observed in a given population are uniformly measured above a fixed $\Sigma_d$ (the same from cloud to cloud), the population should trace out the super-linear relation between $\Sigma_{SFR}$ and $\Sigma_c$ in eq. (\ref{eq:sfrelation}).  
Such a relation might also be expected to apply universally if, as has been suggested, there is a universal column density threshold for star formation (i.e. \citealt{ladall10}; \citealt{heiderman}; \citealt{evansHeiderman}; \citealt{clark}).  In this case, star formation would only ever occur uniformly above a fixed $\Sigma_d$.
 
 \subsubsection{a test with local clouds}
  \label{sec:HeidTest}
We can check for consistency with this prediction in the small sample of 20 Solar Neighborhood clouds mapped in dust extinction with a combination of 2MASS NIR and Spitzer mid-IR data, for which estimates of the star formation rate in the dense gas are available, as compiled by \citet{evansHeiderman} and earlier \citet{heiderman}.  These clouds do not span nearly the range in $\Sigma_c$ as the BH14 sample, but \citet{heiderman} have assembled measurements at multiple locations within each cloud, in contour levels of extinction ranging from $A_v$=2 (typically the lowest level) to $A_v\approx$40.   
Thus for each cloud multiple measurements of the DGMF can be examined for consistency with the model.  Within every contour a star formation rate is estimated from YSO counts and the gas mass is measured throughout the same area assuming a standard conversion from dust extinction to hydrogen column density (see \citet{heiderman} for details).  

The left panel of Figure \ref{fig:heidermanSFR} shows the measured dense gas mass fraction vs. $\Sigma_c/\Sigma_d$ for this sample of clouds, confirming overall very good agreement with the prediction from our simple hydrostatic model in Figure \ref{fig:DGMF_model}.  
In this plot, $\Sigma_c$ and $M_c$ are defined by the outermost (lowest) contour in each cloud.  (Note that this mass need not represent the total cloud mass, which is likely much greater given that the cloud boundary may extend to column densities well below $A_v$=2; \citealt{evansHeiderman}.) 
All other contours tabulated by \citet{heiderman} supply measures of the the dense gas $M_d$ and $\Sigma_d$.  For the black points (based on measurements from \citealt{evansHeiderman}), this contour is uniformly set to $A_v$=8 for all clouds.    

Clouds also exhibit very good agreement with the predicted star formation relation in eq. (\ref{eq:sfrelation}) as shown in the right panel of Figure \ref{fig:heidermanSFR}.  Two different types of measurements are shown there.  
The first case (black symbols) takes measurements (presented without uncertainties) from \citet{evansHeiderman} for the star formation rate (and dense gas mass) measured above a fixed $A_v$=8, corresponding to $\Sigma_d$=120 M$_\odot$ pc$^{-2}$.  The cloud area and $\Sigma_c$ are defined at a lower contour, typically $A_v$=2.   
In the second case (open, gray-scale symbols), each contour defined by \citet{heiderman} within a given cloud supplies a unique measure of $\Sigma_c$ (and $M_c$) and $\Sigma_d$ is defined only in the highest contour level, which varies from cloud to cloud.   
The star formation rate within this uppermost, "highest density" contour supplies an estimate of the star formation rate surface density $\Sigma_{SFR}$ associated with each $\Sigma_c$, after dividing by the cloud area associated with $\Sigma_c$.  

On their own, the black symbols populate only a very narrow region in parameter space (and a limited range in $\Sigma_c$), suggesting that the star formation rate surface densities of clouds are only weakly, if at all, related to their gas surface densities, as previously highlighted by \citet{heiderman}.    
However, taken together with the other measurements (open symbols), these clouds do seem to be well-described by the model relation (with power-law index $N$=2, shown in black).  
By populating a larger area of parameter space, all measurements together also clearly exhibit scatter about the best-fit power-law relation, which is to be expected, given that they span a range in $\Sigma_d$.  
Cases in which $\Sigma_d$ is relatively high (light gray) tend to fall below the line, consistent with the smaller intercept implied by eq. (\ref{eq:sfrelation}), whereas measurements with smaller $\Sigma_d$ (black and dark gray) sit above their counterparts.   

Further validation of the model is apparent in consideration of only those measurements for which $\Sigma_d$ is uniform and set to 120 $M_{\odot} pc^{-2}$ (black symbols).  
The best-fit intercept in this case, which provides a measure of $\log{(\epsilon/2\Sigma_d)}$ (see eq. [\ref{eq:sfrelation}]), remarkably yields $\Sigma_d$=100$\pm$20 $M_{\odot} pc^{-2}$ adopting the efficiency measured by \citet{evansHeiderman} $\log{\epsilon}$=-1.61$\pm$0.23. With the intercept fixed to $\log{(\epsilon/2\Sigma_d)}=-4$, the best-fit slope is exactly $N$=2.  

With so few clouds, the best-fitting line with slope fixed to $N$=1 (not shown) also provides an equally good description of the data, but now the intercept implies $\Sigma_d$=1.6$\pm$0.5 $M_{\odot} pc^{-2}$, two orders of magnitude lower than the value $\Sigma_d$=120 $M_{\odot} pc^{-2}$ expected.  

\subsection{Galaxy-scale star formation}
\label{sec:globalSF}
In the previous section, the pressurized cloud model was shown to provide a compelling explanation for the observed trend between $\Sigma_{SFR}$ and $\Sigma_c$ in local MW clouds.  
The cloud relation is, of course, quite different than the Kennicutt-Schmidt (KS) relation measured on sub-kpc and larger scales in galaxies, in which the measured index of the power-law relation is closer to 1-1.4 \citep{kennEvans}.  The emergence of the latter relation might then be explained as a result of beam dilution and cloud-filling factor considerations (i.e. \citealt{leroy08}; \citealt{lada2013}). 
On the other hand, the KS relation on large scales might arise naturally as a consequence of the regulation of cloud properties by the local galactic environment.  This is explored further below, by considering how ISM pressure regulates the cloud surface density.  
\subsubsection{the role of regulation through pressure} 
In the pressurized cloud model, a gradient in the internal density of a cloud is introduced with the application of non-negligible external pressure at the cloud boundary.  
The cloud boundary, by definition, is where the internal and external pressures are equal.  Thus measurements of the total cloud mass and surface density at the cloud edge are directly related to the pressure in the environment of the cloud, i.e. $P_{ext}=\pi/2G\Sigma_c^2$.  

The external cloud pressure, conversely, originates with the thermal, turbulent and density structure of the surrounding medium itself.  At the mid-plane of the disk, hydrostatic pressure $P_m$ in the ISM of nearby galaxies has been empirically related to the molecular-to-atomic gas ratio $R_{mol}$=$\Sigma_{H2}/\Sigma_{HI}$ by \citet{blitzRos}, where $\Sigma_{H2}$ and $\Sigma_{HI}$ are the H$_2$ and HI gas surface density. The relation is specifically a power-law 
\begin{equation}
R_{mol} = \left(\frac{P_m}{P_0}\right)^{\alpha} \label{eq:rmol}
\end{equation}
in which the empirically measured free parameters are $\log{P_0/k_B}$ = 3.5-4.23 cm$^3 K$ and $\alpha$=0.8-0.94 in nearby galaxies (\citealt{blitzRos}; \citealt{leroy08}), $P_0$ is the external pressure in the interstellar medium (ISM) where the molecular fraction is unity and $P_m$ accounts for the weight of the gas and stellar disks measured according to, i.e. \citet{elmegreen}, 
\begin{equation}
P_m=\frac{\pi}{2}G\Sigma_{gas}\left(\Sigma_{gas}+\frac{\sigma_{gas}}{\sigma_*}\Sigma_*\right) \label{eq:hydroP}
\end{equation}
with $\Sigma_{gas}=\Sigma_{H2}+\Sigma_{HI}$, stellar and gas velocity dispersions $\sigma_{gas}$ and $\sigma_*$ and stellar mass surface density $\Sigma_*$.  This empirical relation is regularly implemented in semi-analytical models (\citealt{dutton}; \citealt{lagos}; \citealt{popping}) and numerical simulations (\citealt{murante10}; \citealt{murante15}) of galaxy formation and evolution to capture the formation of molecular hydrogen in the ISM.   

Now, equating $P_m$ with the external pressure on clouds, the empirical relation between $P_m$ and $R_{mol}$ implies that the cloud surface density and $R_{mol}$ are related, i.e. 
\begin{equation}
\Sigma_c=\sqrt{\frac{2}{\pi}\left(\frac{P_m}{G}\right)}\propto\Sigma_{H2}^{1/2\alpha}\approx\Sigma_{H2}^{1/2} \label{eq:simpleCloud}
\end{equation}

Likewise, the DGMF as described by the equilbrium pressurized cloud model can be written
\begin{eqnarray}
 \frac{M_d}{M_c}=\frac{\Sigma_c}{\Sigma_d}&=&\left(\frac{2}{\pi}\right)^{1/2}\frac{1}{\Sigma_d}\sqrt{\frac{P_m}{G}}\\
 &=&\left(\frac{2}{\pi}\right)^{1/2}\frac{1}{\Sigma_d} \left(\frac{\Sigma_{H2}}{\Sigma_{HI}}\right)^{1/2\alpha}\sqrt{\frac{P_0}{G}}.\label{eq:densepredicGlobal}
 \end{eqnarray}
 
With eq. (\ref{eq:sfrelation}) this immediately implies that
\begin{eqnarray}
 \Sigma_{SFR}&=&\epsilon\left(\frac{M_d}{M_c}\right)\frac{\Sigma_{H2}^{1/2\alpha-1}}{\Sigma_{HI}^{1/2\alpha}}\left(\frac{2}{\pi}\right)^{1/2}\sqrt{\frac{P_0}{G}}\Sigma_{H2} \label{eq:sfrelationGlobal1}\\  
&=&\epsilon\frac{2P_0}{\pi G\Sigma_d}\left(\frac{\Sigma_{H2}}{\Sigma_{HI}}\right)^{1/\alpha}. \label{eq:sfrelationGlobal2}
 \end{eqnarray}

\begin{figure}[t]
\begin{centering}
\begin{tabular}{c}
\includegraphics[width=.95\linewidth]{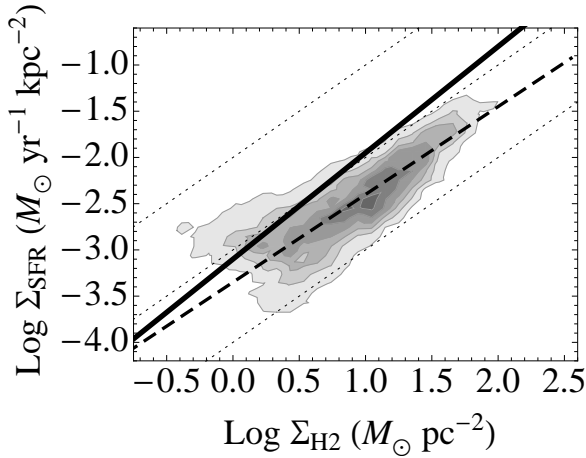}
\end{tabular}
\caption{$\Sigma_{SFR}$ as a function of $\Sigma_{H2}$ in nearby galaxies.  The gray contours show spatially resolved measurements from HERACLES \citep{bigiel}.  The dashed line shows the best-fit relation determined by \citet{leroy2013} given in eq. (\ref{eq:bigiel}) here.  
The solid black line shows the semi-empirical prediction of the pressurized cloud model in eq. (\ref{eq:sfrelationEmpirical}).  
The diagonal dotted lines indicate linear star formation relations, which correspond to star formation that consumes 1\%, 10\%, and 100\% of the gas reservoir in 100 Myr.  These also correspond to lines of constant gas depletion time, 0.1, 1 and 10 Gyr from top to bottom.  
\label{fig:KSplot}
}
\end{centering}
\end{figure}

This expression is in excellent agreement with the molecular star formation relation measured in nearby galaxies by \citet{leroy2013} (and \citealt{bigiel}) (see Figure \ref{fig:KSplot}), 
\begin{equation}
\Sigma_{SFR}=10^{\raisebox{0.1em}{\text{-}}2.35}\Sigma_{H2}^{1.0}. \label{eq:bigiel}
\end{equation} 
Eq. (\ref{eq:sfrelationGlobal2}) specifically yields the semi-empirical prediction 
\begin{equation}
 \Sigma_{SFR}\approx10^{\raisebox{0.1em}{\text{-}}(1.7 \text{--} 2.1)}\Sigma_{H2}^{(1.1\text{--}1.2)}  \label{eq:sfrelationEmpirical}
 \end{equation}
in which the ranges of the indicated exponents are estimated assuming that both the efficiency $\epsilon$ measured by \citet{evansHeiderman} and the dense gas threshold for star formation $\Sigma_d$=160 $M_{\odot} pc^{-2}$ measured by \citet{lada2013} are universal, together with the range of fitted values of $\alpha$ and $P_0$ cited above, and adopting a fixed atomic gas surface density set to the typical threshold value in the range $\Sigma_{HI}$=10-20 $M_{\odot} pc^{-2}$ (e.g. \citealt{leroy08}), which should be applicable in the areas of interest (within normal star-forming disks) but perhaps less so in other environments.  
According to the pressurized cloud model, the relation follows as a consequence of the regulation of both quantities ($\Sigma_{SFR}$ and $\Sigma_{H2}$) by ISM pressure, as discussed further in $\S$ \ref{sec:Pdiscussion}.  

The fact that eq. (\ref{eq:sfrelationEmpirical}) agrees so closely with the empirical star formation relation provides strong, although indirect, support for the pressurized cloud model; observations on large scales in galaxies are consistent with the model based firmly on physics at cloud scales.  
The tight agreement would also tend to favor the idea that $\epsilon$ and $\Sigma_d$ are indeed universal quantities.  
This is explored further below, by examining how the dense gas fraction and the star formation efficiency are predicted to vary with gas surface density.  These predictions, in comparison with observations, provide a more direct test of the model on galactic scales.   

\subsubsection{Predicted trends in dense gas fraction and star formation efficiency across galaxy disks}
\label{sec:predictionsDense}
Consider the simple case in which a region of area $A$ contains $N$ clouds each of area $a=\pi R_c^2$, whereby we write the molecular gas surface density as
\begin{equation}
\Sigma_{H2}=\Sigma_c a \frac{N}{A}+M_{nc}\frac{1}{A}=\Sigma_g a \frac{N}{A}
\end{equation}
where $M_{nc}$ is the mass of molecular material outside clouds and letting $\Sigma_g$ be the molecular surface density on the scale of an individual cloud (accounting for the cloud itself and any surrounding non-cloud molecular material).  Here $N/A$ measures the ``filling factor'' of clouds per unit area.

Similarly, the surface density of dense gas over that area $A$ can be expressed as
\begin{equation}
\Sigma_{dense}=\frac{M_d}{M_c}\Sigma_c a \frac{N}{A}.  
\end{equation}
assuming that the dense gas arises only within clouds.  

Now writing the dense gas fraction over the area $A$ as  
\begin{equation}
f_{dense}=\frac{\Sigma_{dense}}{\Sigma_{H2}}=\frac{M_d}{M_c}\frac{\Sigma_c}{\Sigma_g}\approx\frac{M_d}{M_c} \label{eq:fdense}
\end{equation}
equation (\ref{eq:sfrelationGlobal1}) can be expressed as
\begin{eqnarray}
\Sigma_{SFR}&=&\epsilon_{eff}f_{d}\Sigma_{H2}  \nonumber \\
&=&\epsilon_{eff}\frac{f_{dense}}{\Sigma_c/\Sigma_g}\Sigma_{H2}\label{eq:SFR}
\end{eqnarray}
where the cloud-scale dense gas fraction
\begin{equation}
f_{d}=\frac{M_d}{M_c}=\frac{f_{dense}}{\Sigma_c/\Sigma_g}, 
\end{equation}

the local fraction of molecular gas in cloud form $\Sigma_c/\Sigma_g$ is a factor near unity\footnote{In the standard picture of ISM structure, the molecular phase is entirely in the form of clouds and so the ratio $\Sigma_c/\Sigma_g\approx1$.  It should be noted, though, that recent observations suggest that not all molecular material as traced by CO emission is in cloud form, as mentioned briefly at the end of $\S$ \ref{sec:motivation}.  
This is consistent with the idea of dynamically evolving clouds with short lifetimes and the continuous cycling of molecular material from bound objects to a more diffuse phase, as discussed by \citet{meidtLifetimes}.  Values for $\Sigma_c/\Sigma_g$ might fall below one in this case (as inferred in M51; \citealt{pety}).} and 
\begin{eqnarray}
\epsilon_{eff}&=&\frac{\Sigma_{SFR}}{\Sigma_{dense}}=\epsilon\frac{\Sigma_{c}}{\Sigma_{H2}}\nonumber \\ 
&=&\epsilon\left[\frac{1}{\Sigma_{HI}}\left(\frac{\Sigma_{H2}}{\Sigma_{HI}}\right)^{1/2\alpha-1}\sqrt{\frac{P_0}{G}}\right].  \label{eq:epeff}
\end{eqnarray}

The term in the square brackets in eq. (\ref{eq:epeff}) is related to the filling-factor of molecular gas in cloud form as regulated by pressure (see also eq. \ref{eq:simpleCloud}), i.e. 
\begin{equation}
f_{clouds}=\frac{\Sigma_c}{\Sigma_{H2}}\approx\frac{1}{a}\left(\frac{N}{A}\right)^{-1}
\end{equation}  
when $\Sigma_c a$$>>$$M_{nc}$ ($\Sigma_c/\Sigma_g\approx1$).  Thus, according to eq. (\ref{eq:SFR}) the formation of stars from dense gas first requires the creation of clouds from the molecular medium.   

Notice that both $\epsilon_{eff}$ and $f_{dense}$ share several factors in common, such as $P_0^{1/2}$, and they exhibit roughly inverse dependencies on $\Sigma_{H2}$, leaving a linear relation between $\Sigma_{SFR}$ and $\Sigma_{H2}$ (i.e. eq. [\ref{eq:sfrelationGlobal1}]).  In the simplest case, the pressurized cloud model predicts 
\begin{equation}
f_{dense}\approx10^{\raisebox{0.1em}{\text{-}}(0.6\text{--}0.7)}\left(\frac{\Sigma_{H2}}{\Sigma_{HI}}\right)^{(0.5\text{--}0.6)}  \label{eq:fdensePred}
\end{equation}
and
\begin{equation}
\epsilon_{eff}\approx10^{\raisebox{0.1em}{\text{-}}(1.0\text{--}1.4)}\left(\frac{\Sigma_{H2}}{\Sigma_{HI}}\right)^{\raisebox{0.1em}{\text{-}}(0.4\text{--}0.5)}  \label{eq:epPred}
\end{equation}
where the ranges of the indicated exponents are again estimated 
adopting the empirically measured values for $P_0$, $\alpha$, $\epsilon$ and $\Sigma_d$ in  eqs. (\ref{eq:fdense}) and (\ref{eq:densepredicGlobal}) and eq. (\ref{eq:epeff}) and assuming $\Sigma_c/\Sigma_g=1$.  Thus, in the pressurized cloud model, as the molecular gas surface density (and $\Sigma_{H2}/\Sigma_{HI}$) increases, the measured dense gas $f_{dense}$ should increase.  
But at the same time the efficiency $\epsilon_{eff}$ will decrease in such a way that the star formation efficiency in the molecular gas overall $\Sigma_{SFR}/\Sigma_{H2}$ remains unchanged (and effectively independent of the gas surface density).  

These predictions can be directly tested in nearby galaxies with observations of dense gas tracers like HCN, which, given the appropriate conversion factor, provide measurements of $\Sigma_{dense}$ from which to construct  $\epsilon_{eff}$ and $f_{dense}$ (in combination with existing SFR and molecular gas surface densities).   
It should be emphasized that measured variations in $\epsilon_{eff}$ or  $f_{dense}$ in this context do not necessarily imply that $\epsilon$ or $\Sigma_d$ are non-universal.   
Instead, within the pressurized cloud model, varying $\epsilon_{eff}$ or  $f_{dense}$ would correspond to departures from eqs. (\ref{eq:fdensePred}) and (\ref{eq:epPred}).  Note that, for this reason, disagreement with eqs. (\ref{eq:fdensePred}) and (\ref{eq:epPred}) does not represent invalidation of the pressurized cloud model, which should be considered also in light of quality of the agreement on cloud-scales, such as demonstrated in the previous section.   

\subsubsection{discussion: the impact of local variations in pressure}\label{sec:Pdiscussion}
According to the pressurized cloud model, a linear relation between $\Sigma_{SFR}$ and $\Sigma_{H2}$ follows as a consequence of the regulation of both quantities by ISM pressure.  At the same time that pressure determines the molecular content of the ISM, it also sets the total surface density of clouds, their internal structure and the amount of dense star forming material within them.   

Thus we could argue that the approximately linear relation is not nearly as fundamental as a relation between pressure and star formation.  Indeed, departures from a universal linear relation such as commonly observed between $\Sigma_{SFR}$ and $\Sigma_{H2}$ might be expected to naturally appear as a result of genuine variation in the relation between pressure and $R_{mol}$.  A relation like
\begin{equation}
\Sigma_{SFR}=\frac{\epsilon}{\Sigma_d} \left(\frac{P}{G}\right)
\end{equation} 
(simplifying eq. \ref{eq:sfrelationGlobal2} with eq. \ref{eq:rmol}) 
could be straightforward to implement in modern numerical simulations, where the local ISM pressure on cloud-scales $P$ can be tracked.  

Still, the form of the star formation relation as given by eqs. (\ref{eq:sfrelationGlobal2}) and (\ref{eq:SFR}) may be preferred over a purely pressure-based relation from the observational perspective, given the sensitivity of $P_m$ to the way in which it is estimated (e.g. with or without gas self-gravity; \citealt{blitzRos}) as well as observational limitations to how well the true $P_m$ can be measured.  
Real changes in pressure not captured by hydrostatic equilibrium as expressed in eq. (\ref{eq:hydroP}) might lead the fitted relation $P_m\propto R_{mol}^{1/\alpha}$ to deviate from the actual relation that links these quantities (see also \citealt{feldmannP}).  
This might be constitute a natural source of the scatter in the observed pressure relation traced out by nearby galaxies.  

Changes in pressure would also contribute to the scatter in the star formation relation as expressed by eq. (\ref{eq:sfrelationGlobal2}) measured globally and locally within galaxies.  
For example, the external pressure on clouds can be reduced as a result of gas flows ('dynamical pressure'; \citealt{meidt}).  
This would tend to decouple clouds from their environment.  In this case, with less external pressure to help gravity balance internal turbulent pressure, the result is thought to be a stabilization of clouds against collapse and thus a suppression of star formation \citep{meidt}.  More external pressure, in contrast, would encourage star formation in much the same way that the dissipation of internal turbulent energy within clouds is thought to be a prerequisite for the onset of massive star formation (Hirota et al 2011), i.e. by changing the balance of pressures opposing gravity. 

Such alterations in ISM pressure, as a result of dynamical influences or cloud-scale processes like star formation feedback, would enter eq. (\ref{eq:sfrelationGlobal2}) as variations in $P_0$.  
This quantity encodes the coupling of clouds to their environment, setting the cloud surface density that is linked with a given $\Sigma_{H2}/\Sigma_{HI}$.  Equating $P_0$ with the cloud surface density $\Sigma_{c,0}$ when the molecular-to-atomic ratio is unity (i.e. $\Sigma_{c,0}=\sqrt{P_0/G}$), then according to the pressurized cloud model (equal internal and external pressures at the cloud boundary)
\begin{equation}
\Sigma_c=\Sigma_{c,0}\left(\frac{\Sigma_{H2}}{\Sigma_{HI}}\right)^{1/2\alpha}
\end{equation}
The fitted range in $P_0$ in nearby galaxies corresponds to $\Sigma_{c,0}\approx$ 40 M$_{\odot}$ pc$^{-2}$ but when clouds are decoupled by local processes $\Sigma_{c,0}$ would be decrease. This would in turn reduce $\Sigma_c$, the amount of dense gas in clouds, and the rate of star formation at fixed $\Sigma_{H2}/\Sigma_{HI}$.  
The link between the surface densities of clouds and their local surroundings is at the heart of the star formation relation predicted by the model and should be confirmed, i.e. combining information from cloud-scales and beyond such as now possible in nearby galaxies.  However, this is currently beyond the scope of this paper.  

 \section{Summary/Conclusions }\label{sec:conclusions}
This paper explores the implications of the sensitivity of clouds to external pressure in the disks of normal star-forming galaxies.  
Such sensitivity is implied by the environment-dependence of cloud properties (\citealt{hughesII}; \citealt{colombo2014a}), the frequently observed lack of virialization in clouds (\citealt{keto}; \citealt{heyer09}; \citealt{colombo2014a}) and the near-balance between external and internal cloud pressure recently found in the cloud populations of M51 and other nearby galaxies (\citealt{hughesII}).  
These findings imply the clouds are not decoupled from their surroundings, as commonly assumed.  
 
From within this framework, a new `pressurized' cloud model is developed, in which the application of pressure at the exterior of an effectively isothermal cloud establishes a pressure gradient within the cloud.  
As described by standard equilibrium density profiles, this leads to a strong density contrast from the core to the outside of the cloud.  
The mass of dense gas in the core (i.e the mass above a density threshold) is directly related to the cloud surface density, which is itself inherited from the surrounding environment.  

The internal structure of clouds in the MW, probed by mass measurements at two different densities as compiled by \citet{battistiHeyer},  show strong agreement with the relation between the dense gas mass fraction and cloud surface density predicted by the pressurized cloud model. This presents compelling evidence for the idea that external pressure is important for clouds.  

Further evidence supporting the role of pressure is suggested by the close match between predicted and observed star formation rates both on cloud-scales and 1-kpc scales in galaxies.  
In the pressurized cloud model, star formation is naturally regulated by pressure in the cloud environment:
in a picture in which the star formation rate is set solely by the mass in dense gas within the cloud (as suggested by observations in the MW; \citealt{lada2008}; \citealt{evansHeiderman}) 
then the cloud's surroundings -- which establishes its internal structure and, specifically, the mass in dense gas out of which stars form -- determines the rate at which the cloud forms stars.  
This entails a quadratic dependence of star formation rate surface density on cloud surface density, which is consistent with the trend exhibited by clouds in the Solar neighborhood as observed by \citet{heiderman}.  

Pressure plays a further key role in linking cloud scales with the local environment.  In the pressurized cloud model, the cloud surface density is set by the ambient ISM pressure, which is itself linked to the molecular content of the ISM, according to the empirical relation observed on 1-kpc scales in galaxies \citep{blitzRos}. 
The resulting relation between cloud surface density and the gas surface density on larger scales leads to a linear molecular gas star formation relation, just as parameterized by the Kennicutt-Schmidt relation observed in galaxies.  The pressurized cloud model provides a natural way to connect the two seemingly inconsistent star formation relations.  
The match to both the quadratic cloud-scale relation and the linear 1-kpc scale relation would seem coincidental unless external pressure is important for clouds.  

Existing and future molecular cloud surveys provide ideal test-beds for the `pressurized' cloud model, as they provide an expanded dynamic range in gas densities and pressures and probe a greater diversity of cloud environments than considered here.  
Spatially resolved observations of dense gas tracers on 1-kpc scales and better in nearby galaxies should provide further tests of the model, which predicts specific trends in the dense gas mass fraction and star formation efficiency.  
At present, the model already provides a compelling way to reconcile the star formation in local clouds -- which appeared to be unrelated to global cloud properties -- with the environmental sensitivity of clouds established with observations in external galaxies.  
The result is a consistent basis for describing the environmental regulation of star formation.  

\hspace*{.5in}\\
SEM would like to give special thanks to A. van der Wel for engaged discussion and support during the preparation of this work. Thanks also to the anonymous referee for their careful consideration of the manuscript, to H. Beuther, S. Ragan, A. Hughes and E. Schinnerer for feedback on an early version of the manuscript, to E. Keto and Y. Shirley for useful discussions and to N. Evans and A. Heiderman for notes on data sets appropriate for testing the model.  

\begin{appendix}
\section{Sources of scatter in the DGMF}
\subsection{Observational uncertainty}
Observational uncertainties estimated for the BH14 cloud sample account for assumptions of constant excitation temperature, constant abundance and optical thinness in the $^{13}$CO line used to estimate $M_c$ and $\Sigma_c$ and  
the constant temperature and emissivity assumed for the dust at 1.1 mm tracing the dense gas (see BH14 and references therein).  

Here we consider whether there may also be systematic variation in, i.e., dust temperature and dust-to-gas ratio within the cloud sample that leads to the pattern of scatter in Figure \ref{fig:DGMF_BH14}.  
While the dispersion in the temperatures of BGPS sources (4 K centered on a mean of 14 K) contributes at most 0.16 dex variation in $\Sigma_d$ from its assumed value, we estimate that a factor of 2 (0.3 dex) overestimation in $\Sigma_d$ would require underestimation in the assumed 10 K dust temperature by only 7 K. 
This is not much, considering that cloud interiors could have embedded massive protostars or young clusters that can heat the dust to 40-100 K (BH14).  Thus, the scatter toward high values of $M_d/M_c$ at low $\Sigma_c$ may arise partially with higher than assumed dust temperatures.  

We note that the full 1.5 dex spread from low to high $M_d/M_c$ at $\log{\Sigma_c}$$\sim$2 M$_\odot$pc$^{-2}$ can be explained by interior dust temperatures raising to 40 K.  
However, we do not expect the BGPS sources to consist of such a large number of hot cores as would be required to populate the upper envelope of Figure \ref{fig:DGMF_BH14} (see Eden et al. 2013).  
This would also tend to exclude a preference for hot cores in higher surface density clouds, as might be inferred from the emptiness of the bottom right corner of that Figure.   
At the same time, it seems equally unlikely that the low values of $M_d/M_c$ at low $\Sigma_c$ arise from significant underestimation in $\Sigma_d$ due to overestimation in the dust temperature; the implied cold ($<$3 K) core temperatures are not observed.  
Qualitatively, however, it seems plausible that a higher degree of scatter would appear at low cloud surface density, if the less well-shielded interiors of these clouds are susceptible to larger temperature fluctuations.  

The upper envelope in Figure \ref{fig:DGMF_BH14} could alternatively appear as a result of overestimation in the assumed dust-to-gas ratio (so that the inferred gas masses are higher than in reality).  But if anything, we expect the dust-to-gas ratio to increase, primarily at high density, due to grain growth (e.g. \citealt{Hirashita}).  
In this case, clouds might fall to low $M_d/M_c$ if the assumed dust-to-gas ratio and thus $\Sigma_d$ underestimate their true values.  But then Figure 2 would imply that only the lowest surface density clouds are favorable to grain growth, which seems unlikely.  
\subsection{Cloud size}
\begin{figure}[t]
\begin{centering}
\begin{tabular}{c}
\includegraphics[width=.48\linewidth]{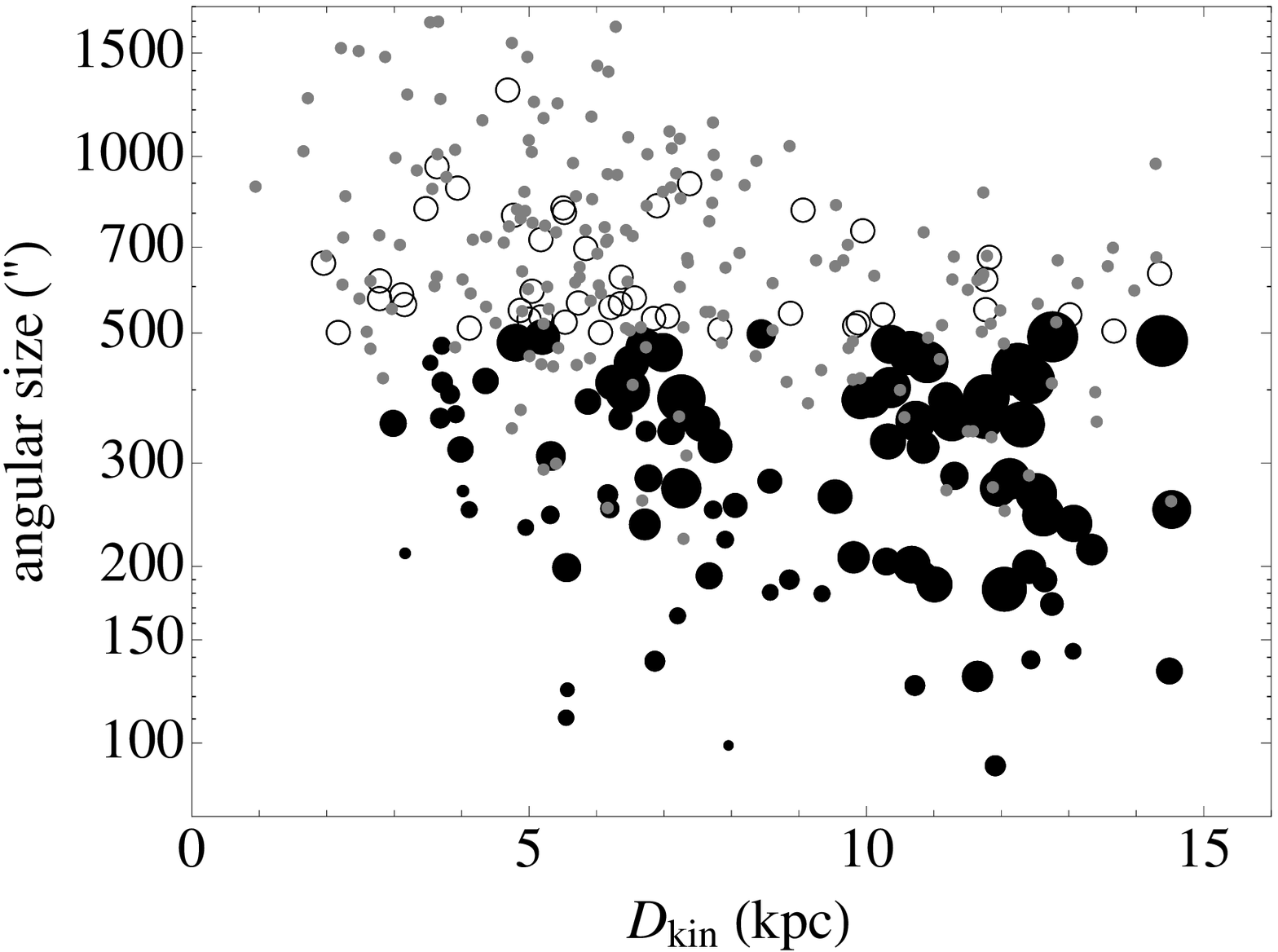}
\end{tabular}
\caption{Plot of the angular sizes of BGPS sources vs. kinematic distance as tabulated by BH14.  
The final subsample of 141 clouds used in the present analysis are shown in black (symbol shape and size are as in Figure \ref{fig:DGMF_BH14}), while objects with multiple BGPS sources along the line of sight, not included in our analysis, as shown as small gray points.  
The latter are on average larger than the studied clouds and are more at risk of underestimated dense gas masses (estimated from median-filtered 1.1mm dust emission; see text).   
\label{fig:angsizeDplot}
}
\end{centering}
\end{figure}
Another potential systematic source of scatter is due to   
the median-filtering applied to the BGPS FIR imaging, which can affect the inferred dust distribution and dense gas masses within clouds depending on their angular size.  Figure \ref{fig:angsizeDplot} shows the angular sizes of the 141 BH14 clouds considered in section \ref{sec:tests}.  
The largest clouds in Figure \ref{fig:DGMF_BH14} tend to have the lowest $M_d/M_c$, as if $M_d$ is systematically underestimated through the filtering process (which removes more FIR flux in objects with larger angular size; BH14; \citealt{aguirre}), and we confirm that clouds fall further down and to the left with increasing physical size.  (The trend is also evident in terms of angular size.)   
But the maximum amount of underestimation should be no more than a factor of 3 for this sub-sample clouds, according to the flux recovery modeled by \citet{aguirre}.  
If we take the measured size of the CO-emitting region as an upper bound on the angular size of the dust-traced high density region at the kinematic distance of the cloud, the measured $M_d$ for the majority of clouds should be well within a factor of 2 of the true value, accounting for no more than $\sim$ 0.25 dex scatter in Figure \ref{fig:DGMF_BH14}.   

Thus it appears that the link between location in Figure \ref{fig:DGMF_BH14} and cloud radius noted above, which also appears as a link with angular size, is not due to observational bias and is physical in origin.  Cloud angular size shows almost not correspondence with kinematic distance in this the sub-sample (Figure \ref{fig:angsizeDplot}).  
Large angular size is thus due to physical size, not merely proximity.\footnote{It should be noted that $M_d/M_c$ appears to vary more strongly with angular size than radius, but this may simply reflect the addition of uncertainty to the measured radius, given the difficulty in determining robust distances to clouds.} 
Moreover, both quantities are arguably fundamentally a measure of cloud mass and are linked to the dense gas mass fraction as an outcome of external pressure, as described by the  pressurized cloud model.  
Only if the cloud mass increases proportionally with radius, as parameterized in eq. (\ref{eq:genratio}), would such a trend emerge; clouds shift down in Figure \ref{fig:DGMF_BH14} as radius increases at the same time as they shift left, due to the radial dependence of $\Sigma_c$. 

\subsection{Deviation from the assumed model}
As considered in $\S$ \ref{sec:DGMFtrends}, variation in the dominant shape of the density profile within the $\Sigma_d$ threshold, from cored (BE) to strictly power-law (SIS), can lead to scatter about the predicted linear relation between $M_d/M_c$ and $\Sigma_c$.  
But there are also additional factors (magnetic fields; non-uniform, scale-dependent turbulence) that can alter cloud equilibrium, potentially invalidating both the BE and the SIS profiles.  
We have also so far ignored other processes (ambipolar diffusion, which moderates the influence of magnetic fields over gravity, turbulent energy dissipation) that lead to fragmentation and collapse within clouds (e.g. \citealt{padoan}; \citealt{hennebelleT}; \citealt{priceBate}; \citealt{VS11}).

\begin{figure}[t]
\begin{centering}
\begin{tabular}{c}
\includegraphics[width=.65\linewidth]{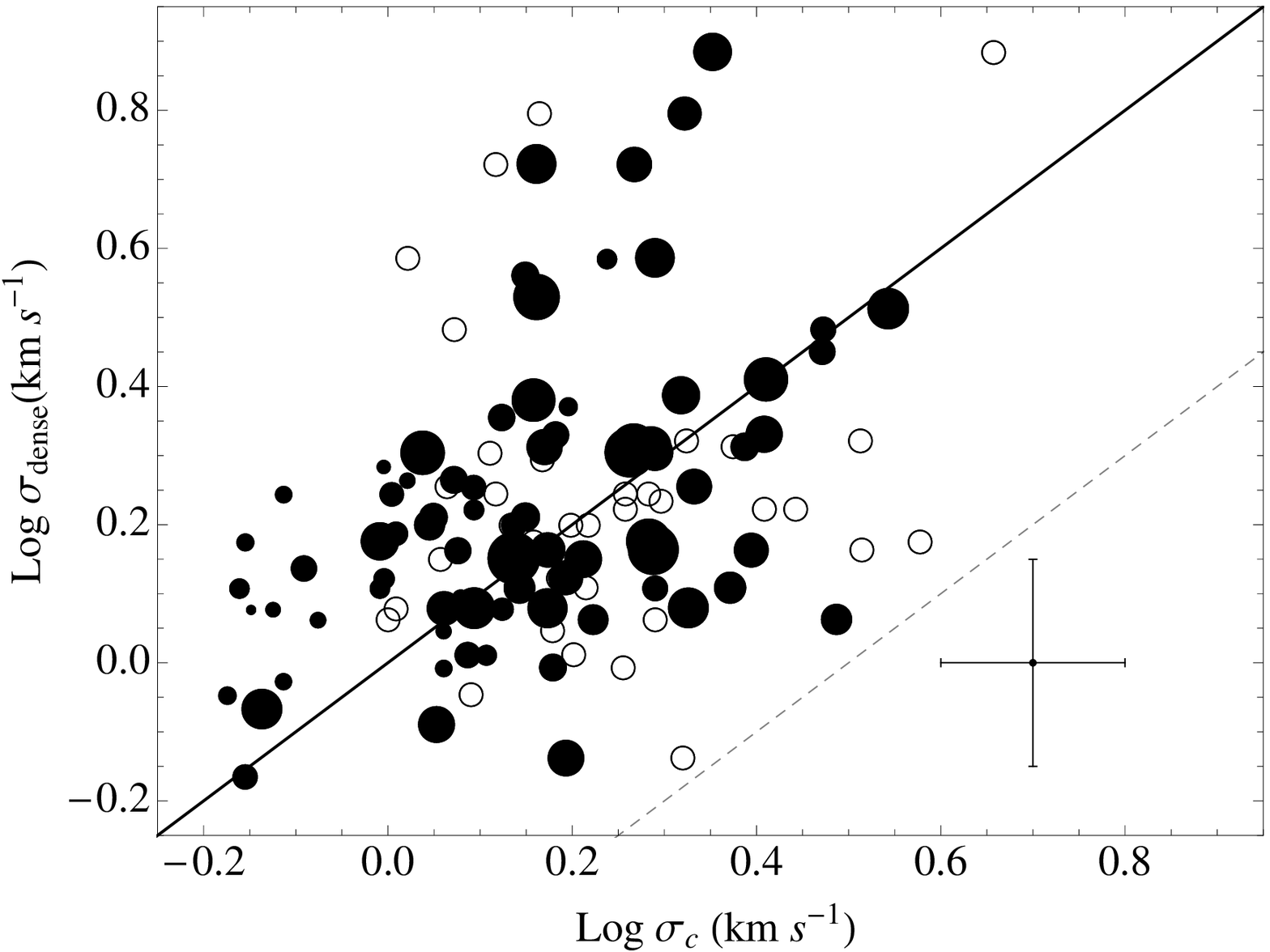}
\end{tabular}
\caption{Velocity dispersion in the dense gas $\sigma_{dense}$ situated within $\sim$ 1 pc in clouds linked to BGPS sources (measured from the HCO$^+$ (3 - 2) line by \citet{shirley} compared to the velocity dispersion $\sigma_c$ measured from the $^{13}$CO line by BH14 over tens of pcs.  
The solid black line shows the one-to-one relation.  The black dotted line indicates the trend between $\sigma_{dense}$ and $\sigma_c$ expected from the size-linewidth relation $\sigma\propto R^{1/2}$ across 1-10pc scales.  
A representative error bar in the bottom right corner shows tabulated 1-sigma uncertainties in $\sigma_{dense}$ and $\sigma_{c}$ that are on average 0.3 km s$^{-1}$ and 0.2 km s$^{-1}$, respectively.  Symbol shapes and sizes are as in Figure \ref{fig:DGMF_BH14}.  
\label{fig:sigplot}
}
\end{centering}
\end{figure}

Based on the overall agreement between our prediction and the clouds in Figure \ref{fig:DGMF_BH14}, though, we would argue that these additional processes are higher-order effects on top of the simple model of external pressure, which serves as the principal basis for the build-up of dense gas across the range in spatial scales considered here.  
In this context the hydrostatic isothermal cloud model can provide a good (if coarse) description since variations in internal velocity dispersion can be largely ignored.  Such variations follow from the turbulent cascade as captured by the size-line width relation across cloud-to-core scales.  
However, between the edge of the cloud and the relatively shallow depths probed by the present threshold $\Sigma_d$, little variation in the velocity dispersion is expected.  
For the BH14 sample, we estimate that the high density gas is typically situated within 0.1-1 $R_c$ (and on average 0.5$R_c$), using the tabulated mass and assuming a constant density $\Sigma_d$=200 M$_\odot$ pc$^{-2}$ throughout the high density region.  
(This provides an upper bound on the effective radius $R_d$ of the high density region in each cloud; the actual area within the $\Sigma_d$ threshold is not tabulated by BH14.)  
According to the \citet{myers} model of turbulent velocity dispersion within clouds, the cloud velocity dispersion radii 0.1-1 pc is very little reduced from its value at the largest scale (the cloud edge).  
Observed line profiles in the dense gas at the cloud centers probed by HCO$^+$ confirm this impression, as shown in Figure \ref{fig:sigplot}.  
For the 114 clouds in our sample with dense gas line widths measured in the HCO$^+$ (3 - 2) line by \citet{shirley}, the line width is comparable to the value measured throughout the entire cloud by BH14 from the $^{13}$CO line.  (Note, though, that unresolved bulk flows tracking infall or outflows may also contribute to the observed linewidth, in addition to turbulence; Shirley et al. 2013).

Although the velocity dispersion may very well decrease to its thermal value at the smallest scales within the cloud (thus presumably impacting the organization of the gas at the very highest densities), a constant velocity dispersion is a valid approximation for the cloud as a whole, since the bulk of the cloud mass is situated between $R_d$ and the cloud edge.  
The SIS and BE profiles should thus provide a coarse description for the development of the internal structure of clouds, though with strong deviations when more of the dense gas (i.e. a large fraction of the cloud mass) is contained at lower $\sigma$.   
\end{appendix}


\end{document}